\documentclass[aps,prx,reprint,superscriptaddress,twocolumn]{revtex4-2}

\usepackage[english]{babel}

\usepackage{amsmath,amssymb,bbm,mathrsfs,mathtools,braket,color,%
  graphicx,comment,textcomp,amsfonts,dsfont,lettrine,units,comment}
\usepackage[colorlinks,citecolor=blue,urlcolor=blue]{hyperref}

\usepackage{algorithm} 
\usepackage{algpseudocode}
\usepackage{array}
\usepackage{hhline}
\usepackage{braket}

\makeatletter
\newcommand*{\transpose}{%
  {\mathpalette\@transpose{}}%
}
\newcommand*{\@transpose}[2]{%
  \raisebox{\depth}{$\m@th#1\intercal$}%
}
\makeatother

\newcommand{\tr}{\mathop{\mathrm{tr}}}

\makeatletter
\newsavebox{\@brx}
\newcommand{\llangle}[1][]{\savebox{\@brx}{\(\m@th{#1\langle}\)}%
  \mathopen{\copy\@brx\kern-0.5\wd\@brx\usebox{\@brx}}}
\newcommand{\rrangle}[1][]{\savebox{\@brx}{\(\m@th{#1\rangle}\)}%
  \mathclose{\copy\@brx\kern-0.5\wd\@brx\usebox{\@brx}}}
\makeatother

\newcommand{\iqoqi}{\affiliation{Institute for Quantum Optics and Quantum Information of the Austrian Academy of Sciences, Innsbruck, Austria}}
\newcommand{\qci}{\affiliation{Institute for Theoretical Physics, University of Innsbruck, Innsbruck, Austria}}

\newcommand{\eqcontr}{\thanks{L.P. and T.O. contributed equally to this work.}}

\begin{document}

\title{Characterization and Verification of Trotterized Digital Quantum Simulation \\ via Hamiltonian and Liouvillian Learning}

\author{Lorenzo Pastori} \eqcontr \qci \iqoqi
\author{Tobias Olsacher} \eqcontr \qci \iqoqi
\author{Christian Kokail} \qci \iqoqi
\author{Peter Zoller} \qci \iqoqi

\date{\today}

\begin{abstract}
The goal of digital quantum simulation is to approximate the dynamics of a given target Hamiltonian via a sequence of quantum gates, a procedure known as Trotterization. The quality of this approximation can be controlled by the so called Trotter step, that governs the number of required quantum gates per unit simulation time. The stroboscopic dynamics generated by Trotterization is effectively described by a time-independent Hamiltonian, referred to as the Floquet Hamiltonian. 
In this work, we propose Floquet Hamiltonian learning to reconstruct the experimentally realized Floquet Hamiltonian order-by-order in the Trotter step. This procedure is efficient, i.e., it requires a number of measurements that scales polynomially in the system size, and can be readily implemented in state-of-the-art experiments. With numerical examples, we propose several applications of our method in the context of verification of quantum devices: from the characterization of the distinct sources of errors in digital quantum simulators to determining the optimal operating regime of the device. We show that our protocol provides the basis for feedback-loop design and calibration of new types of quantum gates. Furthermore it can be extended to the case of non-unitary dynamics and used to learn Floquet Liouvillians, thereby offering a way of characterizing the dissipative processes present in NISQ quantum devices.
\end{abstract}

\maketitle

\section{Introduction} \label{sec:introduction}

Quantum simulation aims at `solving' complex quantum many-body problems efficiently and with controlled errors on quantum devices~\cite{NAP25613,Altman2021}.
The development of quantum simulators is presently considered to be one of the most promising near term goals in quantum technologies, where problems of practical relevance from condensed matter physics~\cite{Lewenstein2012}
to high energy physics~\cite{Aidelsburger2022}
or quantum chemistry~\cite{McArdle2020,Cerezo2021} can be addressed in regimes potentially beyond the abilities of classical computations. Quantum simulation can be implemented as analog or digital quantum simulators. In analog quantum simulation a many-body Hamiltonian of interest finds a natural implementation on a given quantum device. Examples are ultracold bosonic and fermionic atoms in optical lattices as Hubbard models~\cite{Gross2017,Mazurenko2017,Nichols2019,Vijayan2020,Sun2021}, or spin models with trapped ions~\cite{Blatt2012,Monroe2021}, Rydberg tweezer arrays~\cite{Weimer2010,Browaeys2020,Ebadi2021,Scholl2021,Semeghini2021}, and superconducting qubits~\cite{Houck2012,Kjaergaard2020}. While analog simulators naturally scale to large particle numbers, there is limited flexibility realizing complex Hamiltonians required, e.g., in high energy physics and quantum chemistry. In contrast, in digital quantum simulation (DQS) the time-evolution is `programmed' as a sequence of Trotter steps represented by quantum gates~\cite{Lloyd1996}, experimentally realized in several platforms such as trapped ions~\cite{Lanyon2011,Barreiro2011,Martinez2016,Seetharam2021}, Rydberg atoms~\cite{Bluvstein2021}, or superconducting circuits~\cite{Barends2015,Salathe2015,Langford2016}. While in the future fault tolerant and scalable quantum computers might be available to perform DQS, we are still in an era of NISQ devices, i.e., noisy intermediate scale quantum devices, where the elementary quantum operations are performed as analog quantum gates without (the possibility of) error correction. 

A central challenge faced by analog and digital quantum simulation on NISQ devices is thus the characterization and verification of the quantum device, i.e., a demonstration of its proper functioning as a prerequisite for quantitative predictions (see Refs.~\cite{Eisert2020,Carrasco2021} for recent reviews on quantum verification). This is particularly relevant in regimes where classical simulations become intractable, which induces the need to develop efficient protocols for characterization of quantum simulation experiments that scale to large particle numbers. 

Below we address this problem in the context of Trotterized DQS by utilizing recently developed techniques of Hamiltonian learning, which we employ to perform an efficient process tomography of {\em  Trotter blocks} in DQS.

\section{Background and Motivation}
\begin{figure}
	\centering  
	\includegraphics[width=\linewidth]{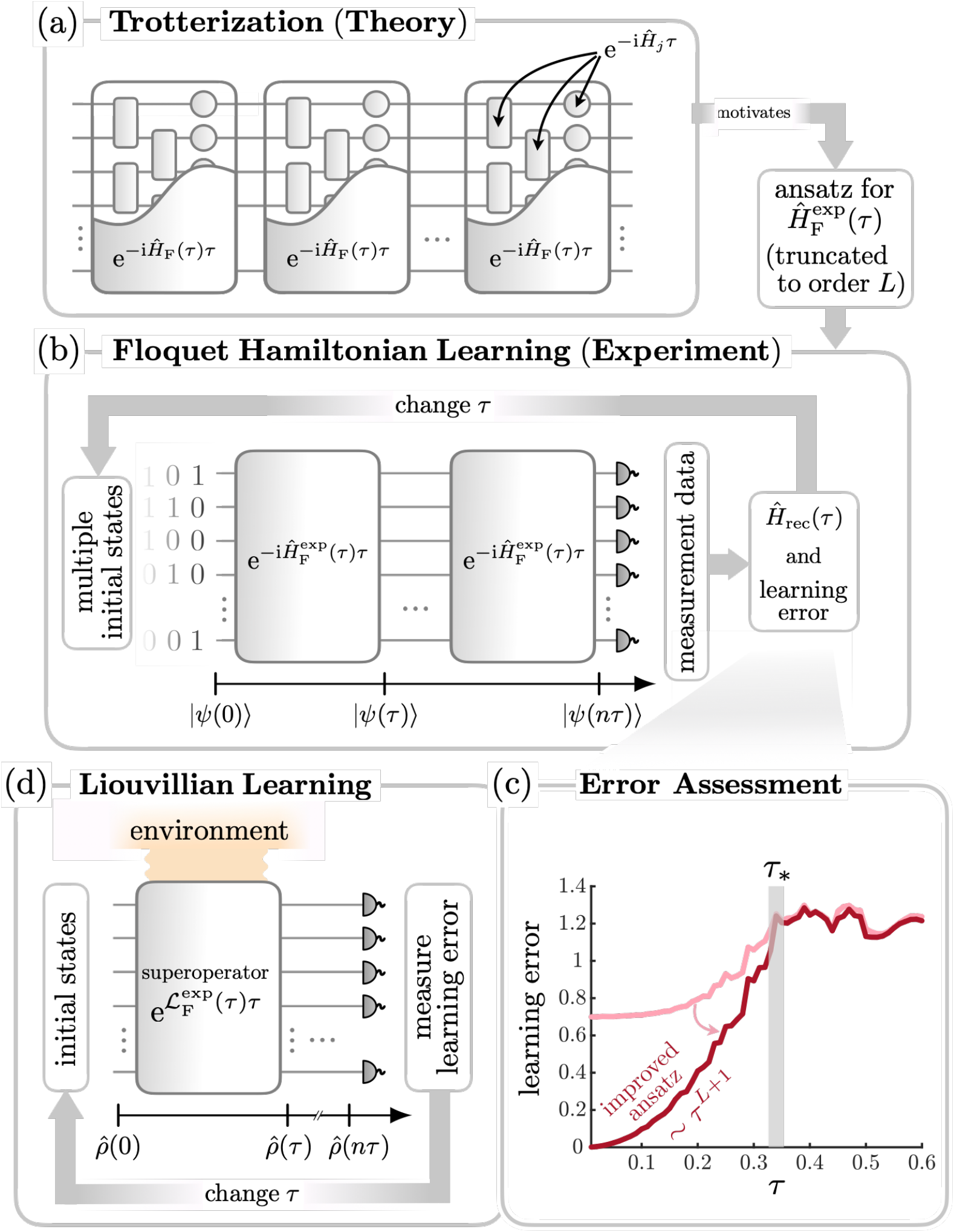}
	\caption{\textbf{Process Tomography of Trotter blocks.} (a) Representation of the Trotter blocks \eqref{eq:trotterblock} which Trotterized DQS aims at implementing. The generator of each Trotter block is the FH $\hat{H}_\mathrm{F}(\tau) \tau$ (multiplied by the Trotter step), defined via Eq.~\eqref{eq:generator}. (b) Trotter blocks implemented in experiments are affected by control errors: they are treated as `black boxes' to be characterized by our protocol. The protocol aims at learning the FH $\hat{H}_{\mathrm{F}}^{{\rm exp}}(\tau)$ from measurements performed on a set of initial states, after repeated application of the Trotter blocks. The measurements are used to fit an ansatz for a truncation of $\hat{H}_{\mathrm{F}}^{{\rm exp}}(\tau)$ to a given order $L$ in $\tau$. The procedure is repeated for several values of $\tau$. (c) The corresponding learning error is measured as a function of $\tau$: if it scales as $\tau^{L+1}$ (for $\tau<\tau_*$), then the reconstruction was successful. The data presented in (c) are obtained from the protocol applied to a DQS of the Heisenberg model Eqs.~\eqref{eq:HeisenbergTarget}-\eqref{eq:HeisenbergTrotteriz}, for learning up to order $L=1$ of the FH. (d) The procedure can be generalized to learning Floquet Liouvillians $\mathcal{L}_{\mathrm{F}}^{{\rm exp}}(\tau)$ in presence of dissipation (see Sec.~\ref{sec:LiouvillianLearning}).}
	\label{fig:illustration}
	\end{figure}

In this section we provide background and motivation for our approach. We start with a brief review of Trotterized DQS and its relation to Floquet systems, and elaborate on motivation and techniques for Floquet Hamiltonian learning as efficient process tomography of Trotter blocks.

\subsection{Trotterized Digital Quantum Simulation}

DQS approximates the time-evolution operator $\mathrm{e}^{-\mathrm{i}\hat{H}t}$
generated by a time-independent many-body Hamiltonian $\hat{H}$ with quantum gate
operations available on a NISQ device. This can be achieved via Trotterization according to a Suzuki-Trotter decomposition~\cite{Suzuki1976,Trotter1959}, as explained below.
We are interested in DQS of generic many-body Hamiltonians with a \emph{local}
structure~\cite{Li2020}, i.e., we write
\begin{align}
    \hat{H}=\sum_{j}\hat{H}_{j}\,,\label{eq:h}
\end{align}
with the Hamiltonian components $\hat{H}_{j}$ being few-body operators. This structure implies that $\hat{H}$ depends on a number of parameters that scales only polynomially in the system size. We emphasize that this notion of locality includes also Hamiltonians with long-range, few-body interactions.
The time-evolution for a sufficiently small
time step $\tau$ is carried out via Trotter blocks $\hat{U}_{\tau}$.
In their simplest form, these Trotter blocks are assembled from the
Hamiltonian components $\hat{H}_{j}$ \footnote{More generally, a Trotter decomposition will be built from a set of
analog resource Hamiltonians $\{\hat{H}_{R}^{(k)}\}_k$ available on
a given NISQ device. These generate a gate set $\{\mathrm{e}^{-\mathrm{i}\hat{H}_{R}^{(k)}t}\}_k$,
which is used to implement the individual components of the Trotter
circuit (\ref{eq:trotterblock}) (see for instance Sec.~\ref{subsec:HeisenbergDisorder}
for a concrete example).} according to
\begin{align} 
    \hat{U}_{\tau}=\prod_j \mathrm{e}^{-\mathrm{i} \hat{H}_j \tau}\,\,,
    \label{eq:trotterblock}
\end{align}
as illustrated in Fig.~\ref{fig:illustration}(a).
The overall coherent time-evolution $\mathrm{e}^{-\mathrm{i}\hat{H}t}$
at times $t=n\tau$ ($n=1,2,\ldots$) is thus approximated by a sequence
of Trotter blocks $\hat{U}_{\tau}^{n}$. This becomes exact when
$\tau\rightarrow0$, i.e., $\mathrm{e}^{-\mathrm{i}\hat{H}t}=\lim_{n\rightarrow\infty}\hat{U}_{t/n}^{n}$. While the Trotter block in Eq.~\eqref{eq:trotterblock} approximates $\hat{H}$ up to deviations of order $\tau$, we refer to \cite{Wiebe2010,Chen2022} for decompositions with higher order Trotter errors. The following discussion, and the methods developed in the present work, will apply to any Trotterization.

Trotterized DQS has the structure of a Floquet problem~\cite{Else2020,Heyl2019,Sieberer2019,Kargi2021},
i.e., a periodically driven many-body problem~\cite{Bukov2015} with stroboscopic time
evolution generated by $\hat{U}_{\tau}$. Writing the Trotter block as
\begin{align}
    \hat{U}_{\tau}=\mathrm{e}^{-\mathrm{i}\hat{H}_{\mathrm{F}}(\tau)\tau}\,\, ,
    \label{eq:generator}
\end{align}
we can identify $\hat{H}_\mathrm{F}(\tau)$ with the Floquet Hamiltonian (FH) of the driven system.

The stroboscopic time-evolution generated by DQS is thus $\hat{U}_{\tau}^{n} = \mathrm{e}^{-\mathrm{i}\hat{H}_{\mathrm{F}}(\tau)n\tau}$, to be compared with the targeted time-evolution $\mathrm{e}^{-\mathrm{i}\hat{H}n\tau}$. Therefore we can interpret $\hat{H}_\mathrm{F}(\tau)$ as the physical or `actual' Hamiltonian implemented for a given Trotter step $\tau$ on the quantum device.

The FH plays a key role in our experimental protocol for efficient process tomography of a Trotter block below, and thus we will briefly elaborate on its properties. 
For sufficiently small $\tau$ we can write the FH as a Magnus expansion~\cite{Bukov2015}
\begin{align}
    \hat{H}_{\mathrm{F}}(\tau)=\sum_{l=0}^{\infty}\hat{\Omega}_l\,\tau^l\,\,,
    \label{eq:hf}
\end{align}
and for reference we write out the first few terms explicitly for the Trotter decomposition  Eq.~\eqref{eq:trotterblock} in Appendix \ref{app:Magnus}.
Here $\hat{\Omega}_{0} = \hat{H}$ is the target Hamiltonian
of the DQS, while higher order terms $\hat{\Omega}_{l}$
($l>0$) represent \textit{Trotter errors}, written as perturbation to the original Hamiltonian $\hat H$.

In Eq.~(\ref{eq:hf}) we distinguish the regime where the Magnus expansion converges, i.e., the regime where DQS is a controlled approximation to the target Hamiltonian. On the other hand,
divergence of the series leads to proliferation of Trotter errors at a critical step size $\tau_{*}$ \cite{Moan2008}. The existence of this \emph{Trotter threshold}
\cite{Heyl2019,Sieberer2019,Kargi2021} is intrinsically connected
to the implemented dynamics turning quantum chaotic for $\tau>\tau_{*}$,
where the spectrum of $\hat{U}_{\tau}$ exhibits universal properties
described by random matrix theory. This is a phenomenon extensively studied with simple model systems as the kicked top ~\cite{Haake2018,Sieberer2019,Olsacher2022}, and in the context of periodically driven many-body systems~\cite{Haake2018,DAlessio2014,Regnault2016,Kos2018}.

\subsection{Process Tomography of Trotter Blocks via Floquet Hamiltonian and Liouvillian Learning}

Below we will be interested in efficient process tomography of Trotter blocks in an experimental setting, where the Trotter block is considered as a `black box' to be characterized in an experiment. To this end we will apply techniques of Hamiltonian learning to infer the FH parametrizing the Trotter block.

Hamiltonian learning (HL) techniques were originally developed as protocols to infer local Hamiltonians of closed systems from steady states~\cite{Garrison2018,Qi2019,Chertkov2018,Bairey2019,Evans2019,Hou2020,Kokail2021,Anshu2021}, as well as methods for reconstructing Hamiltonians governing the time-evolution in quench dynamics~\cite{Wiebe2014,Wang2015,Li2020,Bienias2021,Hangleiter2021,Zubida2021,Yu2022} from measurements performed on the quantum device. The central assumption underlying HL is that typically implemented many-body Hamiltonians have an operator structure consisting of a small number of terms that scales polynomial in the system size.
The prototypical HL protocol starts with an ansatz for the operator content of the Hamiltonian to be learned, and infers a best fit from comparison with experimental data. These protocols also have a built-in toolbox
for error assessment in the learned Hamiltonian, allowing
for a systematic exploration by enlarging the ansatz to
include missing terms. 

In the present context, we note that the target Hamiltonian Eq.~\eqref{eq:h} was assumed to have a local structure, and this property is inherited by any truncation of the FH \eqref{eq:hf} for $\tau<\tau_{*}$ (see Appendix \ref{app:Magnus}). This allows HL developed for quench dynamics to be applied to learn the FH implemented by a DQS.

In contrast to the ideal case of Eq.~\eqref{eq:hf}, in implementing Trotter blocks in an experiment there will be errors, e.g., control errors so that the experimentally realized Trotter block $\hat{U}_{\tau}^{{\rm exp}}$
will differ from the ideal $\hat{U}_{\tau}$. More generally, quantum gates will suffer from decoherence, i.e., the experimental Trotter block must be described by open system dynamics as a Kraus map, instead of a unitary operator. This is schematically shown in Fig.~\ref{fig:illustration}(b), where the Trotter blocks are treated as `black boxes' to be characterized.

Starting with the unitary case, our goal is thus to learn the experimental FH $\hat{H}_\mathrm{F}^\mathrm{exp}(\tau)$ defined via
\begin{align}
 \hat{U}_{\tau}^{{\rm exp}} =\mathrm{e}^{-\mathrm{i}\hat{H}_{\mathrm{F}}^{{\rm exp}}(\tau)\tau}\,\,.
 \label{eq:generator-1}
\end{align}
This is achieved by inferring the operator content of
\begin{align}
    \hat{H}_{\mathrm{F}}^{{\rm exp}}(\tau)=\sum_{l=0}^{\infty}\hat{\Omega}_{l}^{{\rm exp}}\,\tau^{l} \,\,,
    \label{eq:hf-1}
\end{align}
\emph{order-by-order} in $\tau$ (for $\tau<\tau_\ast$). Specifically, we formulate an ansatz for a truncation of Eq.~\eqref{eq:hf-1} to a chosen order $\tau^{L}$, and fit it to measurement data obtained from the output states $\ket{\psi(n \tau)} = (\hat{U}_{\tau}^\mathrm{exp})^n \ket{\psi_0}$ for chosen initial states $\ket{\psi_0}$, where experiments are performed for various $n=1,2,\ldots$ and $\tau$ \footnote{We make the reasonable assumption that control errors in experimentally repeated Trotter blocks are identical, i.e. $\hat{U}_{\tau}^\mathrm{exp}$ is independent of $n$.}. 
As discussed in detail in Sec.~\ref{sec:HLinDQS}, the behavior of the learning error as a function of $\tau$ reveals whether a $L$-order truncation of $\hat{H}_{\mathrm{F}}^{{\rm exp}}(\tau)$ has been learned: a learning error scaling with $\tau^{L+1}$ confirms a successful reconstruction, while the scaling with a lower power suggests a systematic extension of the ansatz by additional terms.
This is illustrated in Fig.~\ref{fig:illustration}(c), where we show numerical data obtained from the application of our protocol to a DQS of the Heisenberg model (see Sec.~\ref{subsec:HeisenbergDisorder}): an incomplete ansatz leads to a plateau of the learning error (pink curve) while scaling (red curve) is observed if and only if all the terms up to order $\tau^{L}$ are included.
The number of samples required for such a reconstruction scales only polynomially with the system size, as long as $\tau<\tau_{*}$. Furthermore, monitoring the learning error as a function of $\tau$ provides a mechanism to detect the Trotter threshold $\tau_\ast$, as the point at which learning $\hat{H}_{\mathrm{F}}^{{\rm exp}}(\tau)$ fails: for $\tau>\tau_{*}$, an ansatz consisting of a polynomial number of terms generally cannot approximate the FH, which is reflected in a sharp increase of the learning error (marked by the grey vertical line in Fig.~\ref{fig:illustration}(c)).

Note that by learning the \textit{operator content} of  $\hat{H}_{\mathrm{F}}^{{\rm exp}}(\tau)$, and comparing it with the target Hamiltonian $\hat H$, both control and Trotter errors are expressed as perturbations to the original (target) Hamiltonian $\hat H$, and are thus directly amenable to physical interpretation; i.e., by learning the operator structure of these perturbations we expect direct insight regarding their physical effects in simulating quantum phases and quench dynamics \footnote{An example is provided by `learning' the experimental FH in DQS of lattice gauge theories with gauge variant and invariant Hamiltonian error terms~\cite{GonzalezCuadra2022}.}.
We contrast this to measuring Trotter block or quantum gate fidelities by comparing the experimental $\hat{U}_{\tau}^{{\rm exp}}$ with the targeted $\mathrm{e}^{-\mathrm{i}\hat H \tau}$, which provides only a global assessment of errors \footnote{We note that in general process tomography scales exponentially with the system size.}. 
Finally, in an experiment the FH learning protocol can be run in a quantum feedback loop to finetune parameters as a way of engineering desired Trotter blocks, which will even work in a regime where the quantum device cannot be accurately calibrated \cite{Marciniak2022}.

While the above discussion has focused on coherent evolution, these ideas are readily extended to include dissipative processes (as schematically shown in Fig.~\ref{fig:illustration}(d)). Assuming that the dissipative errors can be modelled as a quantum Markov process, i.e., described by a master equation, we can learn in a similar way the operator structure of the corresponding Floquet Liouvillian $\mathcal{L}^{\rm exp}_\mathrm{F}(\tau)$, which also admits a Magnus expansion in powers of $\tau$ in analogy to Eq.~\eqref{eq:hf-1} (see Eq.~\eqref{eq:LLMagnusExp} in Sec.~\ref{sec:LiouvillianLearning}).

The remainder of this  paper is organized as follows. In Sec.~\ref{sec:HLinDQS} we introduce the technical details of our procedure for process tomography of Trotter blocks including order-by-order HL of Floquet Hamiltonians. Sec.~\ref{sec:NumResults} is dedicated to several possible applications of this scheme: We illustrate the characterization of Trotter errors in Sec.~\ref{subsec:HeisenbergDisorder}, the detection of imperfect implementation of Trotter blocks in Sec~\ref{subsec:AdaptiveErrors}, a verification of a DQS of the lattice Schwinger model in Sec.~\ref{subsec:Schwinger} as well as the design of many-body gates in Sec.~\ref{subsec:ClusterModel}. Finally in Sec.~\ref{sec:LiouvillianLearning} we extend our order-by-order learning scheme to the characterization of dissipative processes in the context of Floquet Liouvillian learning.

\section{Order-by-order Learning of the Floquet Hamiltonian $\hat{H}^\mathrm{exp}_{\rm F}(\tau)$}
\label{sec:HLinDQS}
In this section we describe the protocol for process tomography of Trotter blocks in DQS, based on quantifying the individual terms in the experimental Floquet Hamiltonian Eq.~(\ref{eq:hf-1}) order-by-order in $\tau$. We refer to this as Floquet Hamiltonian learning (FHL). Before going into the details, we provide a brief overview of Hamiltonian tomography in quench dynamics, which serves as the basis for our protocol. While in the following we focus on Hamiltonian learning, i.e., assuming unitary dynamics, we will include dissipative processes in the context of Floquet Liouvillian learning in later sections.

\subsection{Hamiltonian learning in quench dynamics}
Hamiltonian learning techniques, as developed in the context of non-equilibrium many-body dynamics, form an important sub-routine of our protocol. In this paper we exploit different methods for learning local Hamiltonians from projective measurements as introduced in Ref.~\cite{Li2020} and Ref.~\cite{Bairey2019}. We start by discussing the scheme of Ref.~\cite{Li2020}, which outlines a method to recover the structure of a time-independent Hamiltonian $\hat{H}$ that governs the unitary time-evolution in a set of quantum quenches $\ket{\psi_i(t)} = \mathrm{e}^{-\mathrm{i} \hat{H} t} \ket{ \psi_i(0) } $, by measuring local observables on the output states $\ket{\psi_i(t)}$. Here, $\{ \ket{\psi_i(0)} \}_{i=1}^{N_\text{con}}$ denote a set of (in general) arbitrarily chosen initial states (see below). 
The basic idea is to reconstruct the Hamiltonian $\hat{H}$ implemented on the quantum device from a parametrized ansatz Hamiltonian
\begin{align}\label{eq:ansatz}
\hat{H}(\boldsymbol{c}) = \sum_{\hat{h}_j \in \mathcal{A}} c_j \hat{h}_j
\end{align}
which is composed of $N_{\mathcal{A}}$ few-body operators $\hat{h}_j$ chosen from a set $\mathcal{A}$ whose number scales polynomially with the system size. 
Within this ansatz class, the goal is now to find optimal parameters $\boldsymbol{c}^\text{rec}$ which yield the best possible approximation to the physically implemented Hamiltonian $\hat{H}$ from the condition of energy conservation. In particular, for a time-independent Hamiltonian we have the conditions $\braket{\hat{H}}_{i, t} = \braket{\hat{H}}_{i, 0} $ for all $i$, in which $\braket{\hat{H}}_{i, t} = \langle\psi_i(t)|\hat{H}|\psi_i(t)\rangle$. As proposed in Ref.~\cite{Li2020}, such an optimal approximation for the ansatz Hamiltonian can be determined by minimizing the energy differences between initial and final states over the parameters $\boldsymbol{c}$, i.e.
\begin{align} \label{eq:argmin}
\boldsymbol{c}^{\text{rec}} = \underset{\boldsymbol{c}, \ \Vert \boldsymbol{c} \Vert = 1}{\text{arg}\,\text{min}} \,\sum_{i=1}^{N_\text{con}} \left[ \braket{\hat{H}(\boldsymbol{c})}_{i, t} - \braket{\hat{H}(\boldsymbol{c})}_{i, 0} \right]^2\,\,.
\end{align}
This is a quadratically constrained problem where we optimize under the condition that $\boldsymbol{c}^\mathsf{T}\boldsymbol{c} = 1$. Note that the parameters are learned only up to an overall scale, which can be determined separately via time-resolved measurement of a single observable (see App.~\ref{app:OverallScale}).
Every quench experiment starting from an initial state $\ket{\psi_i(0)}$ provides an independent constraint on the coefficients $\boldsymbol{c}$, implying that in general $N_\text{con} > N_{\mathcal{A}}$ constraints are required for a unique reconstruction. 
The cost function (\ref{eq:argmin}) has the advantage that it can be readily expressed in matrix form using the expansion Eq.~(\ref{eq:ansatz}). In particular we can rewrite Eq.~(\ref{eq:argmin}) as
\begin{align}
    \boldsymbol{c}^{\text{rec}} = \underset{\boldsymbol{c}, \ \Vert \boldsymbol{c} \Vert = 1}{\text{arg}\,\text{min}} \; \boldsymbol{c}^{\mathsf{T}} M^{\mathsf{T}} M \boldsymbol{c} \,\,,
\end{align}
where the matrix elements $M_{i,j} = \langle\hat{h}_j\rangle_{i,t}  -  \langle\hat{h}_j\rangle_{i,0}$ correspond to local observables to be measured on the output states $\ket{\psi_i(t)}$. Minimizing (\ref{eq:argmin}) thus amounts to identifying $\boldsymbol{c}^{\text{rec}}$ with the right singular vector corresponding to the smallest singular value $\lambda_1$ of $M$~\cite{Li2020}
\begin{align}
    \lambda_1  = \Vert M \boldsymbol{c}^\text{rec} \Vert \,\,.
    \label{eq:lambda1}
\end{align}
Here a non-zero value of $\lambda_1$ is caused by the ansatz set $\mathcal{A}$ being insufficient to account for the operator content of the implemented $\hat{H}$. Thus, we refer to $\lambda_1$ as the \emph{learning error}, as its value becomes an indicator for the quality of the reconstructed $\boldsymbol{c}^{\text{rec}}$. The learning error will be the central quantity for order-by-order learning of $\hat H_{\rm F}^\mathrm{exp}(\tau)$.

\subsection{Protocol for learning the operator content of the Floquet Hamiltonian order-by-order}
\label{sec:oboFHL}

We now turn to the discussion of learning the individual orders $\hat{\Omega}^{\mathrm{exp}}_l$ of the Floquet Hamiltonian $\hat{H}^{\mathrm{exp}}_{\mathrm{F}}(\tau) = \sum_{l = 0}^{\infty} \hat{\Omega}^{\mathrm{exp}}_l \tau^l$, that is the generator of an experimental Trotter block $\hat{U}^{\mathrm{exp}}_\tau$. 
We will start from an ansatz for a truncation of $\hat H^{\mathrm{exp}}_{\rm F}(\tau)$ in the form of Eq.~\eqref{eq:ansatz}, and attempt the reconstruction of the parameters $\boldsymbol{c}$ using the method above.

The central idea of our approach is to repeat the reconstruction for several values of $\tau$ with the chosen ansatz components $\{\hat{h}_j\}$, and to measure the learning error $\lambda_1$ as a function of the Trotter step $\tau$. The behavior of $\lambda_1(\tau)$ with $\tau$ is used to rigorously assess whether the given ansatz captures the operator content of all $\hat{\Omega}^{\mathrm{exp}}_l$ up to the chosen truncation order $L$.
Specifically, the scaling $\lambda_1(\tau) \sim \tau^{L+1}$ ensures that the chosen ansatz captures all components of $\hat{H}^{\mathrm{exp}}_{\rm F}(\tau)$ up to order $\tau^{L}$, thus \emph{verifying} the reconstruction.
The key element enabling this rigorous error assessment is the fact that a scaling $\lambda_1(\tau) \sim \tau^{L+1}$ implies an equivalent scaling for the distance between the implemented $\hat{H}^{\mathrm{exp}}_{\rm F}(\tau)$ and the reconstructed Hamiltonian $\hat{H}_{\rm rec}(\tau)\equiv\hat{H}(\boldsymbol{c}^{\text{rec}}(\tau))$, i.e.,
\begin{align} \label{eq:ScalingRelation}
    \lambda_1(\tau) \sim \tau^{L+1} \Leftrightarrow \Vert \hat{H}_{\rm rec}(\tau) - \hat{H}^{\mathrm{exp}}_{\mathrm{F}}(\tau) \Vert \sim \tau^{L+1} \,\,,
\end{align}
which we prove in Appendix \ref{app:ScalingRelation}.
Importantly, this result enables us to distinguish between Trotter errors and control errors that will appear in an experimental setting. For example, if an ansatz $\mathcal{A}$ encompassing the target Hamiltonian $\hat{H}$ of the DQS leads to a constant value of $\lambda_1(\tau)$, we can conclude that additional terms, corresponding to experimental imperfections, need to be added to $\mathcal{A}$. Once these terms have been added and $\lambda_1(\tau)\sim\tau$, one can further improve the ansatz by adding (nested) commutators to learn higher order terms $\hat{\Omega}^{\mathrm{exp}}_{l}$, thereby achieving an order-by-order learning of $\hat{H}^{\mathrm{exp}}_{\rm F}(\tau)$.

Our protocol thus comprises the following steps:
\begin{itemize}
    \item[(1)] Expand the ansatz Hamiltonian $\hat{H}(\boldsymbol{c})$ in a local operator basis according to Eq.~(\ref{eq:ansatz}). The operator basis $\mathcal{A} = \mathcal{A}_0 \cup \mathcal{A}_1 \cup \dots \cup \mathcal{A}_L$ is first of all motivated from the structure of the terms appearing in the (theoretical) Magnus expansion, i.e., $\mathcal{A}_l$ contains the $l$-nested commutators appearing in $\hat{\Omega}_l$ (see App.~\ref{app:Magnus}). In an experimental setting, $\mathcal{A}$ should then include additional terms whose presence we want to test.
    
    \item[(2)] For a given $\tau$, choose a set of initial states $\{ \ket{\psi_i(0)} \}$ and simulation times $\{ n \tau \}$.
    
    \item[(3)] Construct the constraint matrix $M(\tau)$, by measuring the matrix elements 
    \begin{equation} 
        M_{i,j}(\tau) = \braket{ \hat{h}_j}_{i,0} - \braket{ \hat{h}_j}_{i,n\tau} \,\,,
        \label{eq:Melements}
    \end{equation}
    with $\braket{ \hat{h}_j}_{i,n\tau} = \braket{ \psi_i(n\tau) \vert \hat{h}_j \vert \psi_i(n\tau)}$ and $\ket{\psi_i(n\tau)}=(\hat{U}^{\mathrm{exp}}_{\tau})^n\ket{\psi_i(0)}$.
    
    \item[(4)] Reconstruct the optimal parameters $\boldsymbol{c}^\text{rec}$ by calculating the smallest singular value $\lambda_1(\tau)$ and the corresponding right singular vector of $M(\tau)$.
    
    \item[(5)] Repeat the steps (1)-(4) for different Trotter step sizes $\tau$ and study the behavior of $\lambda_1(\tau)$ as a function of $\tau$. If $\lambda_1(\tau)$ is constant for $\tau<\tau_\ast$ the set $\mathcal{A}$ needs to be complemented with additional operators until a scaling $\lambda_1(\tau) \sim \tau^{L+1}$ is observed. If $\lambda_1(\tau) \sim \tau^{L+1}$, $\hat{H}_{\text{rec}}(\tau)$ captures all orders of $\hat{H}^{\mathrm{exp}}_{\rm F}(\tau)$ up to order $L$.
\end{itemize}

The Magnus expansion for $\hat{H}^{\mathrm{exp}}_{\rm F}(\tau)$ is guaranteed to converge only up to a critical value of $\tau$~\cite{Moan2008}, that is connected to the breakdown of Trotterization at the Trotter threshold at $\tau_*$~\cite{Heyl2019,Sieberer2019,Kargi2021}. 
In general, for $\tau > \tau_*$ the Floquet Hamiltonian $\hat{H}^{\mathrm{exp}}_{\rm F}(\tau)$ does not exhibit a local structure, meaning that process tomography based on a local ansatz will fail which is reflected in a sharp increase of $\lambda_1$. Therefore our protocol provides an experimentally feasible way of detecting the Trotter threshold in DQS, and thus probing the regime of validity of the Trotter approximation (see Sec.~\ref{subsec:HeisenbergDisorder}).

In practice, the matrix elements in $M(\tau)$ will be noisy due to the fact that they are estimated from a finite number of projective measurements. As we will show, the effect of a finite number of samples on $\lambda_1$ can be analytically estimated, and does not spoil the validity of the protocol. We will discuss this in more detail in the examples presented in the next section.

Among the several types of imperfections, static deviations from the desired resources/target Hamiltonians can be detected with our protocol by including the corresponding terms in the ansatz set $\mathcal{A}$. Moreover, Eq.~(\ref{eq:argmin}) is agnostic to state preparation errors as long as sufficiently many independent constraints are provided.

Note that while the presentation in Sec.~\ref{sec:HLinDQS} puts emphasis on learning the Hamiltonian in the context of coherent dynamics, the entire procedure can be extended to dissipative dynamics which we present in Sec.~\ref{sec:LiouvillianLearning}. In such a more general Liouvillian learning procedure the Hamiltonian as well as the dissipative terms are parametrized by an ansatz in Lindblad form and the corresponding parameters are reconstructed from the minimization of an experimentally measurable cost function in analogy to Sec.~\ref{sec:HLinDQS}. Similar to Eq.~\eqref{eq:ScalingRelation} the scaling of the minimum of the cost function can be used to verify the quality of the reconstruction.

\section{Floquet Hamiltonian Learning in DQS: Numerical Examples} \label{sec:NumResults}

In this section, we present numerical results on some proposed applications of Floquet Hamiltonian learning. We do so in the context of a DQS of several experimentally relevant systems, i.e., the XXZ/Heisenberg model, the lattice Schwinger model and the 1D cluster model. All implemented Trotter gate sequences are built from the following set of resources
\begin{equation} \label{eq:resources}
	    \mathcal{R} = \{\mathrm{e}^{-\mathrm{i} J\hat{\sigma}^x_i\hat{\sigma}^x_j},\hat{R}^{(j)}_{\boldsymbol{n}}(\theta)\} \,\,,
\end{equation} 
for any pair of qubits $i,j$ and with tunable parameters $J$, $\theta$ and rotation axis $\boldsymbol{n}$, with $\hat{R}^{(j)}_{\boldsymbol{n}}(\theta)=\mathrm{e}^{-\mathrm{i}\theta\boldsymbol{n}\cdot\hat{\boldsymbol{\sigma}}_j}$. Here and in the following, $\hat{\sigma}^{\mu}_j$ ($\mu=x,y,z$) denote Pauli operators acting on the $j$-th qubit.

\subsection{Disordered XXZ-Heisenberg Model} \label{subsec:HeisenbergDisorder}

\begin{figure*}
	\centering  
	\includegraphics[width=0.95\linewidth]{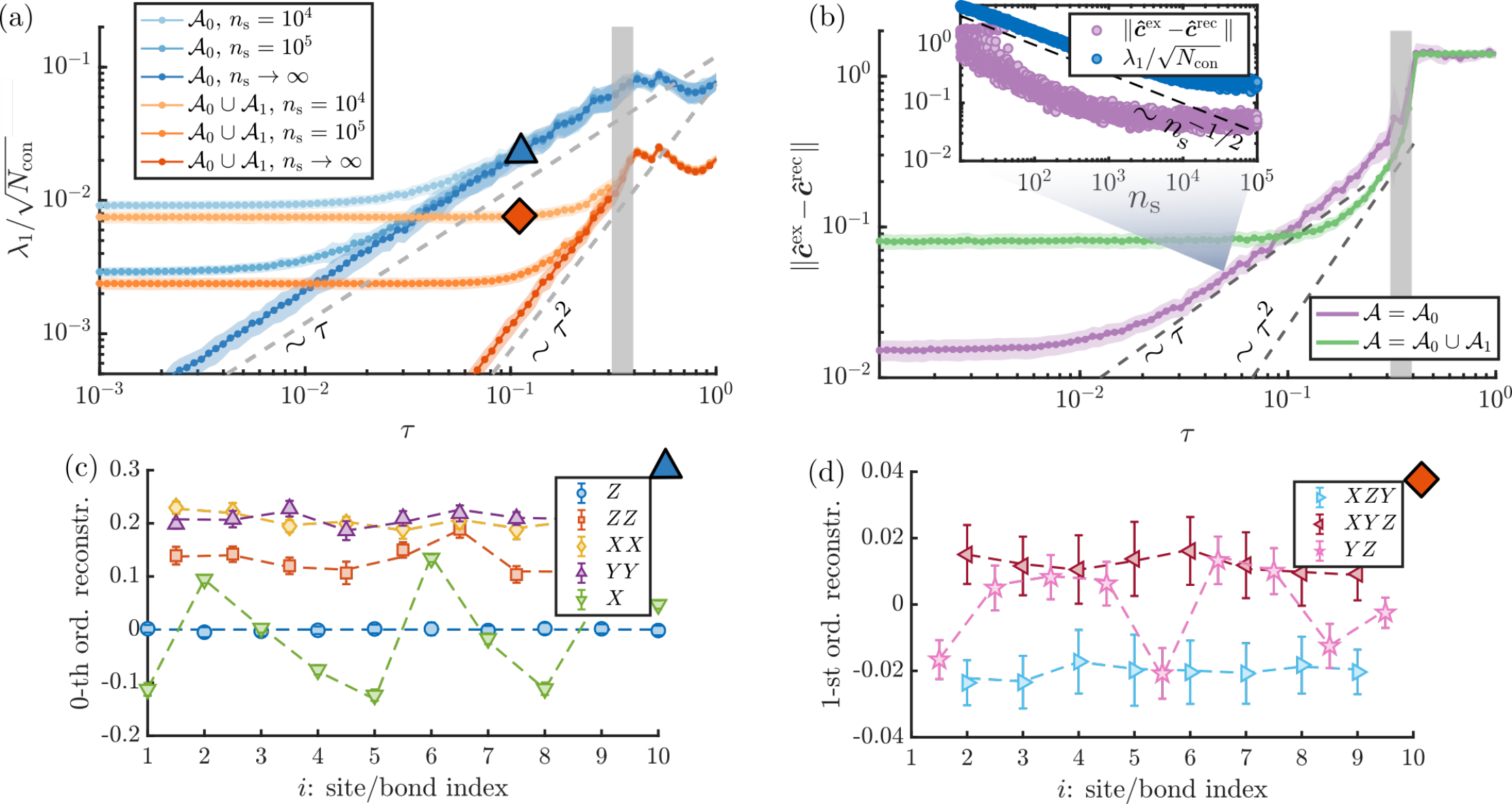}
	\caption{{\bf{Floquet Hamiltonian learning in DQS of the XXZ-Heisenberg model}}. (a) Scaling of $\lambda_1/\sqrt{N_{\mathrm{con}}}$ with $\tau$ for a zeroth (blue) and first (orange) order ansatz, and different measurement budgets $n_{\mathrm{s}}$. The plateau value at small $\tau$ can be predicted using Eq.~\eqref{eq:Elambda}. (b) Reconstruction error $\Vert\hat{\boldsymbol{c}}^{\mathrm{ex}}-\hat{\boldsymbol{c}}^{\mathrm{rec}}\Vert$ vs.~$\tau$ for $n_{\mathrm{s}}=10^4$, for zeroth (purple) and first (green) order ansatz. Here, $\hat{\boldsymbol{c}}^{\mathrm{rec}}$ denotes the normalized reconstructed coefficient vector (similarly for $\hat{\boldsymbol{c}}^{\mathrm{ex}}$). The scaling of $\lambda_1/\sqrt{N_{\mathrm{con}}}$ is reflected in that of $\Vert\hat{\boldsymbol{c}}^{\mathrm{ex}}-\hat{\boldsymbol{c}}^{\mathrm{rec}}\Vert$, as predicted by Eq.~\eqref{eq:ScalingRelation}. The inset shows the scaling of $\lambda_1/\sqrt{N_{\mathrm{con}}}$ and $\Vert\hat{\boldsymbol{c}}^{\mathrm{ex}}-\hat{\boldsymbol{c}}^{\mathrm{rec}}\Vert$ with $n_{\mathrm{s}}$ for a fixed $\tau\approx 0.05$. In both (a) and (b) the vertical grey line denotes the value of $\tau$ at which Trotter errors proliferate. (c) Reconstructed Hamiltonian coefficients to zeroth order for $\tau=0.098$, showing perfect agreement with the exact values (dashed lines). The coefficients of terms that are absent from the Hamiltonian are reconstructed to be zero (blue data points). (d) Reconstructed Hamiltonian coefficients to first order for the same value of $\tau$. Here the coefficients of the normalized $\hat{\boldsymbol{c}}^{\mathrm{rec}}$ are shown, while in App.~\ref{app:OverallScale} we show how to extract their overall scale. The results are obtained for a system of $N=10$ spins, with $J^{x/y}_j=1+\delta_j^{x/y}$ with $\delta_j^{x/y}$ randomly distributed in $[-0.15,0.15]$, $J^{z}_j=0.7+\delta_j^{z}$ with $\delta_j^{z}\in[-0.25,0.25]$, and $B^x_j\in[-0.75,0.75]$. We use $55$ initial product states in random bases, each evolved to $6$ different final times up to a total simulation time of $16$. The error bars stem from measurement noise and from different initial states realizations.}
	\label{fig:Heisenberg}
\end{figure*}

As a first example we consider process tomography of Trotter blocks for a spin-1/2 XXZ-model. In particular we study the scaling relation for $\lambda_1$ in Eq.~\eqref{eq:ScalingRelation} and show how one can detect the breakdown of Trotterization from the behavior of $\lambda_1$ as a function of $\tau$.
In a second step, we discuss the effects that a finite measurement budget has on $\lambda_1$. Finally, we show that our protocol can be employed to learn disorder patterns in quantum many-body systems, which is important in the context of, e.g., many-body localization~\cite{Abanin2019}.

The target Hamiltonian considered here is
\begin{equation} \label{eq:HeisenbergTarget}
    \hat{H}=\sum_{j=1}^{N-1}\sum_{\mu=x,y,z}J^{\mu}_j\hat{\sigma}^{\mu}_{j}\hat{\sigma}^{\mu}_{j+1}+\sum_{j=1}^{N}B^x_j\hat{\sigma}^{x}_{j} \,\,.
\end{equation}
A possible Trotterization for this model reads
\begin{equation} \label{eq:HeisenbergTrotteriz}
    \hat{U}_{\tau}=\hat{U}^{ZZ}_{\tau}\,\hat{U}^{YY}_{\tau}\,\hat{U}^{XX}_{\tau}\,\hat{U}^{X}_{\tau} \,\,,
\end{equation}
where the individual terms can be generated from the resources \eqref{eq:resources} as
\begin{align*}
    & \hat{U}^{XX}_{\tau}=\prod_j\mathrm{e}^{-\mathrm{i}\tau J^x_j\hat{\sigma}^x_j\hat{\sigma}^x_{j+1}} \,\,,\\
    & \hat{U}^{YY}_{\tau}=\hat{R}_z(\tfrac{\pi}{4})\prod_j\mathrm{e}^{-\mathrm{i}\tau J^y_j\hat{\sigma}^x_j\hat{\sigma}^x_{j+1}}\hat{R}_z(-\tfrac{\pi}{4}) \,\,,\\
    & \hat{U}^{ZZ}_{\tau}=\hat{R}_y(\tfrac{\pi}{4})\prod_j\mathrm{e}^{-\mathrm{i}\tau J^z_j\hat{\sigma}^x_j\hat{\sigma}^x_{j+1}}\hat{R}_y(-\tfrac{\pi}{4}) \,\,,\\
    & \hat{U}^{X}_{\tau}=\prod_j\hat{R}^{(j)}_x(B^x_j\tau) \,\,,
\end{align*}
with $\hat{R}_{\mu}(\theta)=\prod_j\hat{R}^{(j)}_{\mu}(\theta)=\prod_j\mathrm{e}^{-\mathrm{i}\theta\hat{\sigma}^{\mu}_j}$ being rotations on all qubits ($\mu=x,y,z$). 
In this section, for FHL we choose a zeroth order ansatz
\begin{align*}
    \mathcal{A}_0\,&=\{Z,ZZ,XX,YY,X\}\\
    & \equiv\{\hat{\sigma}^z_j,\hat{\sigma}^z_j\hat{\sigma}^z_{j+1},\hat{\sigma}^x_j\hat{\sigma}^x_{j+1},\hat{\sigma}^y_j\hat{\sigma}^y_{j+1},\hat{\sigma}^x_j\}_{j=1,...,N} \,\,,
\end{align*}
and a first order ansatz
\begin{align*}
    \mathcal{A}_0\cup\mathcal{A}_1=\mathcal{A}_0\,\cup\,& \{Y,ZY,YZ,XY,YX,XZY,YZX, \\
    & XYZ,YXZ,ZXY,ZYX\} \,\,,
\end{align*}
comprising all the terms in $\hat{\Omega}_0$ and $\hat{\Omega}_1$ in the implemented Floquet Hamiltonian. The case of an `incomplete' ansatz set will be treated in Sec.~\ref{subsec:AdaptiveErrors}.

In Fig.~\ref{fig:Heisenberg}(a) we show the behavior of $\lambda_1$ as a function of $\tau$ for $\mathcal{A}_0$ (blue data points) and $\mathcal{A}_0\cup\mathcal{A}_1$ (orange data points), plotted for different measurement budgets, represented here by the number $n_{\mathrm{s}}$ of measurements used for the estimation of each of the expectation values that appear in the constraint matix $M$. Let us first focus on the data for $n_{\mathrm{s}}\to\infty$. There, for $\tau$ smaller than a critical value $\tau_*$ marking the Trotter threshold, we can observe scaling of $\lambda_1$ with $\tau$. A zeroth order ansatz leads to a scaling $\lambda_1\sim \tau$, confirming the successful reconstruction of all zeroth order terms in $\hat{\Omega}_0$, and the presence of (Trotter) errors proportional to $\tau$. Similarly, a first order ansatz leads to a scaling $\lambda_1\sim \tau^2$ indicating the presence of Trotter errors proportional to $\tau^2$. Analogous scaling can be observed in the \emph{parameter distance} $\Vert\boldsymbol{c}^{\mathrm{ex}}(\tau)-\boldsymbol{c}^{\mathrm{rec}}(\tau)\Vert$ between the vectors of reconstructed ($\boldsymbol{c}^{\mathrm{rec}}(\tau)$) and exact ($\boldsymbol{c}^{\mathrm{ex}}(\tau)$) Hamiltonian coefficients, shown in Fig.~\ref{fig:Heisenberg}(b): this is a consequence of the scaling relation \eqref{eq:ScalingRelation}.

Sharp threshold behavior of $\lambda_1$ at $\tau \approx \tau_*$ signals the Trotter threshold, marked by a vertical grey line in Figs.~\ref{fig:Heisenberg}(a)-(b). For $\tau>\tau_*$, the FH 
cannot be approximated by a local ansatz, which is signaled by the absence of scaling of $\lambda_1(\tau)$ and $\Vert\boldsymbol{c}^{\mathrm{ex}}(\tau)-\boldsymbol{c}^{\mathrm{rec}}(\tau)\Vert$. In this sense the behavior of $\lambda_1(\tau)$ can be used as an experimentally measurable signature for the proliferation of Trotter errors. For a deeper discussion of FHL in the context of the Trotter threshold and the transition to quantum chaos, we refer the reader to Appendix \ref{app:TrotterThreshold} and to our recent work on collective spin systems~\cite{Olsacher2022}. 

In contrast to the limit $n_{\mathrm{s}}\to\infty$,  a finite measurement budget leads to a plateau of $\lambda_1$ at small values of $\tau$, as shown in Fig.~\ref{fig:Heisenberg}(a). This is due to the presence of (simulated) measurement noise on the elements of the constraint matrix $M$ defined in Eq.~\eqref{eq:Melements}, that limits the resolution of $\lambda_1$. The value of this plateau for a given measurement budget can be estimated as
\begin{equation} \label{eq:Elambda}
	\lambda_1(\tau\to 0) \lesssim \sqrt{\frac{N_{\mathrm{con}}-N_{\mathcal{A}}+1}{n_{\rm s}}} \,\,,
\end{equation}
as shown in Appendix \ref{app:Elambda}. The quality of the reconstruction in presence of measurement noise, as measured by $\lambda_1$ or the parameter distance, is proportional to $n_{\rm s}^{-1/2}$, as shown in the inset of Fig.~\ref{fig:Heisenberg}(b). The estimate in Eq.~\eqref{eq:Elambda} is valid when errors due to an incomplete ansatz are smaller than the noise introduced by a finite measurement budget. Thus, comparing $\lambda_1(\tau)$ to the above estimate allows one to decide whether the ansatz needs to be complemented with additional terms, as we discuss in Sec.~\ref{subsec:AdaptiveErrors} below.

In Fig.~\ref{fig:Heisenberg}(c)-(d) we show the reconstructed Hamiltonian parameters to zeroth ($\hat{\Omega}^\mathrm{exp}_0$) and first order ($\hat{\Omega}^\mathrm{exp}_1$), respectively, for a given value of the Trotter step $\tau$. There, the spatial dependence of the coefficients is accurately reproduced for all operators appearing in the first orders of the FH 
(the dashed lines show the exact values). This demonstrates the ability of FHL in resolving the spatial structure of the Hamiltonian parameters, thereby learning the disorder pattern characterizing a given many-body system. The overall scale of the coefficients can be efficiently extracted using the method described in App.~\ref{app:OverallScale}.

\subsection{Detection of Static Control Errors} \label{subsec:AdaptiveErrors}

In this section, we show how our protocol can be used to detect, and learn, possible errors affecting the resource gates in a DQS. The errors considered here are of two types: (i) deviations of the resource Hamiltonians from the expected ones, and (ii) static errors in the single-qubit rotations. In the latter case, our protocol also allows for estimating the optimal value of the Trotter step for which the DQS most faithfully reproduces the target Hamiltonian.

To understand the origin of these two types of errors in a specific example we consider the Trotterization for the XXZ/Heisenberg model used in the previous section, which we rewrite here for a translation invariant Hamiltonian as
\begin{equation} \label{eq:HeisenbergTrotterizTranslInv}
\begin{split}
	\hat{U}_{\tau}= &\,\hat{R}_y(\tfrac{\pi}{4})\,\mathrm{e}^{-\mathrm{i}\tau J^z\hat{H}_{XX}}\hat{R}_y(-\tfrac{\pi}{4})\\
	&\times \hat{R}_z(\tfrac{\pi}{4})\,\mathrm{e}^{-\mathrm{i}\tau J^y\hat{H}_{XX}}\hat{R}_z(-\tfrac{\pi}{4})\\
	&\times \mathrm{e}^{-\mathrm{i}\tau J^x\hat{H}_{XX}}\hat{R}_x(B^x\tau) \,\,,
	\end{split}
\end{equation}
with $\hat{H}_{XX}=\sum_j\hat{\sigma}^x_j\hat{\sigma}^x_{j+1}$. The entangling gates $\mathrm{e}^{-\mathrm{i}\tau J^{\mu}\,\hat{H}_{XX}}$ are implemented as the time-evolution with the analog resource Hamiltonian $\hat{H}_{XX}$ over `times' $\tau J^{\mu}$: in realistic implementations, the operator content of $\hat{H}_{XX}$ may slightly deviate from the expected one (i.e., $\sum_j\hat{\sigma}^x_j\hat{\sigma}^x_{j+1}$) by additional unknown terms. We refer to these terms as deviations from the expected resource Hamiltonians. Additionally, a DQS can be affected by errors in the qubit rotations $\hat{R}_{\mu}(\theta)$, resulting in generally small and site-dependent over- or under-rotations. We now address these two cases separately, showing how our protocol can be used to detect and learn the corresponding imperfections.

\subsubsection{Static deviations from expected resource Hamiltonians}

\begin{figure} [t]
	\centering  
	\includegraphics[width=1.0\linewidth]{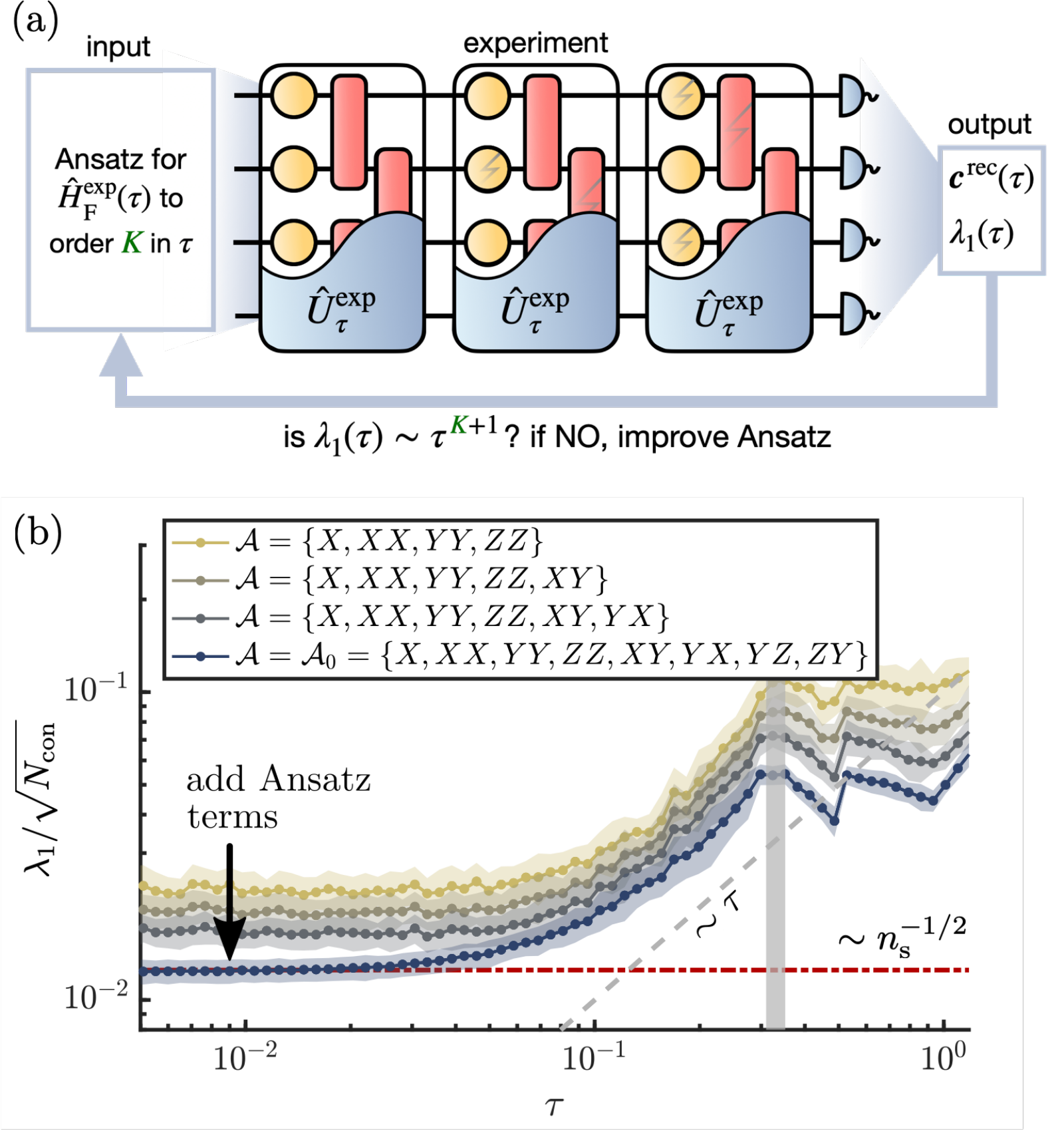}
	\caption{{\bf{Adaptive learning of unknown terms}}. (a) Schematic view of the adaptive learning scheme, aimed at observing scaling \eqref{eq:ScalingRelation} in $\lambda_1(\tau)$. (b) Loglog plot of $\lambda_1(\tau)/\sqrt{N_{\mathrm{con}}}$ at different stages of the adaptive learning scheme, for $n_{\mathrm{s}}=5000$. An incomplete zeroth order ansatz leads to a constant value of $\lambda_1/\sqrt{N_{\mathrm{con}}}$, that is larger than the noise threshold Eq.~\eqref{eq:Elambda} (red dashed line in the figure). In this case add terms to the ansatz until we reach the noise threshold. The blue curve corresponds to the full zeroth order ansatz $\mathcal{A}_0$, with error terms $\hat{V}_{j,j+1}$ in the $XX$ gates having the form $\hat{V}_{j,j+1}=\alpha_j(\hat{\sigma}^x_{j}\hat{\sigma}^z_{j+1} + \hat{\sigma}^z_{j}\hat{\sigma}^x_{j+1})+\alpha^2_j\hat{\sigma}^z_{j}\hat{\sigma}^z_{j+1}$ (with $\alpha_j\approx0.1$). The results are obtained for a system of $N=10$ spins, with uniform couplings $J^x=J^y=1$, $J^z=0.75$, $B^x=0.5$ in Eq.~\eqref{eq:HeisenbergTarget}. We use $55$ initial product states in random bases, each evolved to $6$ different final times up to a total simulation time of $16$. The error bars stem from measurement noise and from different initial states realizations.}
	\label{fig:AdaptiveHeis}
\end{figure}

We consider an imperfect implementation of the $XX$ resource Hamiltonian that appears in Eq.~\eqref{eq:HeisenbergTrotterizTranslInv} of the form
\begin{equation*}
	\hat{H}_{XX}=\sum_j\hat{\sigma}^x_j\hat{\sigma}^x_{j+1} \to \sum_j\big(\hat{\sigma}^x_i\hat{\sigma}^x_{j+1}+\hat{V}_{j,j+1}\big)\,\,,
\end{equation*}
with $\hat{V}_{j,j+1}$ being an unknown two-body operator acting on qubits $j$ and $j+1$. To learn the terms $\hat{V}_{j,j+1}$, we employ FHL with the aim of constructing an ansatz set $\mathcal{A}$ for which scaling $\lambda_1\sim\tau$ is observed down to the \emph{noise threshold} of Eq.~\eqref{eq:Elambda}. We do so by repeatedly adding to $\mathcal{A}$ new operators to be measured, until this condition is met. Errors or unknown terms generated by physical processes typically correspond to few-body operators: this restricts the search for new operators to a space whose dimension scales only polynomially with the system size. 

We illustrate this \emph{adaptive} protocol in Fig.~\ref{fig:AdaptiveHeis}(a) and further apply it to a DQS of the translation invariant XXZ/Heisenberg Hamiltonian Eq.~\eqref{eq:HeisenbergTarget} in Fig.~\ref{fig:AdaptiveHeis}(b).
While an ansatz based on the target Hamiltonian leads to a plateau of $\lambda_1$ at a value which is significantly higher than the noise threshold (the red dashed line in the figure), the addition of ansatz terms encompassing the unknown operators lowers this value. When the ansatz is such that the plateau value of $\lambda_1$ falls below the noise threshold, we can conclude that if additional unknown terms are present, they are small and cannot be resolved with the current measurement budget.

\begin{figure} [t]
	\centering  
	\includegraphics[width=1\linewidth]{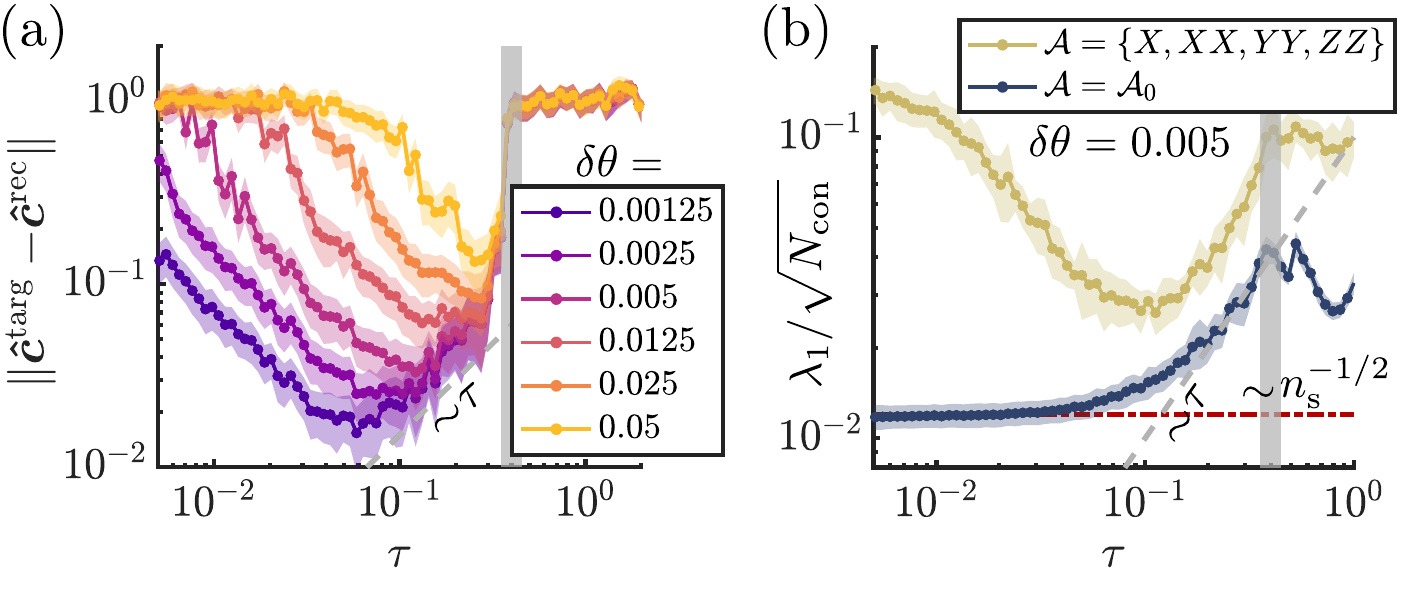}
	\caption{{\bf{Rotation errors and optimal Trotter step}}. (a) Parameter distance $\Vert\hat{\boldsymbol{c}}^{\mathrm{targ}}-\hat{\boldsymbol{c}}^{\mathrm{rec}}\Vert$. Different colors correspond to different strengths $\delta\theta$ of the rotation errors. Here, $\hat{\boldsymbol{c}}^{\mathrm{rec}}$ and $\hat{\boldsymbol{c}}^{\mathrm{targ}}$ are normalized to $1$. The dip in $\Vert\hat{\boldsymbol{c}}^{\mathrm{targ.}}-\hat{\boldsymbol{c}}^{\mathrm{rec}}\Vert$, which can be used to estimate the optimal $\tau$ for the DQS in presence of rotation errors. (b) Loglog plot of $\lambda_1$ vs.~$\tau$ for an ansatz encompassing the target Hamiltonian \eqref{eq:HeisenbergTarget} (yellow data), and a full zeroth order ansatz set $\mathcal{A}_0$ (blue data, which leads to scaling down to the noise threshold. The results are obtained for a system of $N=10$ spins, with target couplings $J^x=J^y=1$, $J^z=0.7$, $B^x=0.75$ in Eq.~\eqref{eq:HeisenbergTarget}. Here $n_{\mathrm{s}}=5000$. We use $55$ initial product states in random bases, each evolved to $6$ different final times up to a total simulation time of $16$. The error bars stem from measurement noise and from different initial states realizations.}
	\label{fig:RotErrsHeis}
\end{figure}

\subsubsection{Rotation errors, and determining the optimal Trotter step}

We consider now the case in which the rotations in Eq.~\eqref{eq:HeisenbergTrotterizTranslInv} are affected by qubit-dependent over- or under-rotations with strength $\delta\theta$, as
\begin{equation*}
	\hat{R}_{\mu}(\theta)\to \hat{R}_{\mu}(\theta+\boldsymbol{\delta\theta})=\prod_j\mathrm{e}^{-\mathrm{i}(\theta+\delta\theta_j)\hat{\sigma}^{\mu}_j}\,\,,
\end{equation*}
with $\delta\theta_j$ uniformly sampled in $[-\delta\theta,\delta\theta]$. Such rotation errors become more important for smaller Trotter steps: the optimal Trotter step at which the DQS most accurately simulates the target Hamiltonian is determined by a tradeoff between rotation errors and Trotter errors, dominant for large $\tau$. Here, we show that such optimal Trotter step can be determined using FHL.

In Fig.~\ref{fig:RotErrsHeis} we show an example of our scheme applied in presence of rotation errors. In Fig.~\ref{fig:RotErrsHeis}(a) we show the parameter distance $\Vert\boldsymbol{c}^{\mathrm{targ}}-\boldsymbol{c}^{\mathrm{rec}}\Vert$ between the reconstructed coefficients and the target ones $\boldsymbol{c}^{\mathrm{targ}}$ (the ones corresponding to the target Hamiltonian), as a measure of the accuracy of the DQS. Different colors correspond to different strengths $\delta\theta$ of the rotation errors. As can be seen in Fig.~\ref{fig:RotErrsHeis}(a), for each $\delta\theta$, there is a clearly visible minimum separating the regimes of dominating Trotter errors (large $\tau$) and dominating rotation errors (small $\tau$): this minimum can be used as an estimate for the optimal Trotter step for the DQS. As shown in Fig.~\ref{fig:RotErrsHeis}(b), the same behavior is observed in the learning error $\lambda_1(\tau)$ for an ansatz encompassing the target Hamiltonian \eqref{eq:HeisenbergTarget}, i.e., $\mathcal{A}=\{X,XX,YY,ZZ\}$ (yellow data). An ansatz capturing all operators generated by the imperfect rotations leads to scaling down to the noise threshold, as expected.

\begin{figure} [t]
	\centering  
	\includegraphics[width=1.0\linewidth]{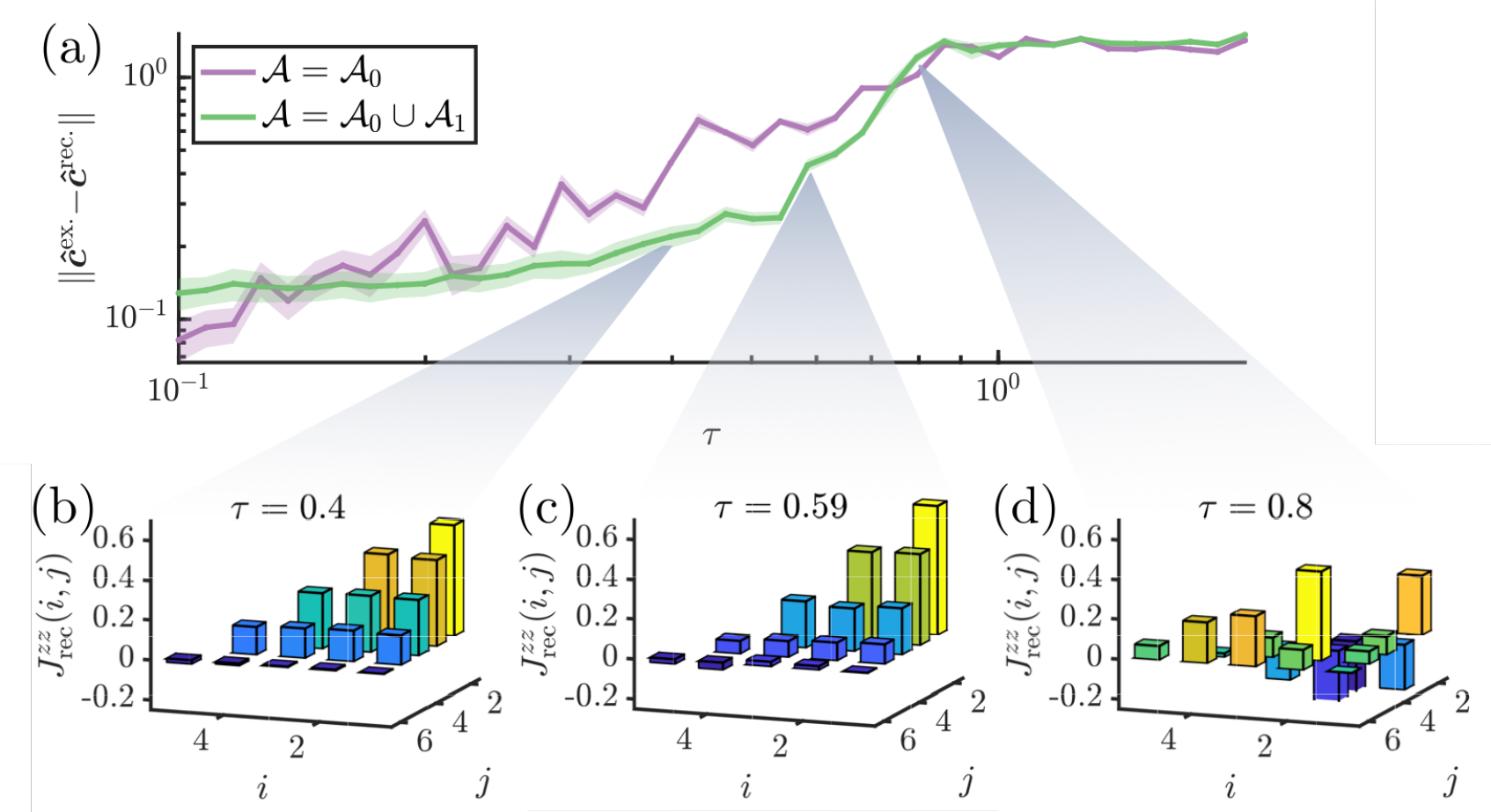}
	\caption{{\bf{FHL for DQS of Schwinger Model}}. Numerical simulation of FHL applied to a DQS of the lattice Schwinger model with $N=6$ qubits. Panel (a) shows the reconstruction error $\Vert \hat{\boldsymbol{c}}^{\mathrm{ex}}-\hat{\boldsymbol{c}}^{\mathrm{rec}}\Vert$ as a function of $\tau$ for ansätze up to 0th and 1st order. The error bars stem from a finite measurement budget per operator $n_{\mathrm{s}}=10^3$. Panels (b)-(d) show the reconstructed $zz$-interactions for several Trotter steps. Here we use time-evolution of $100$ random initial product states for final times $T \in \{2,3 \}$,  yielding a total of $200$ constraints for $\vert \mathcal{A}_0 \cup \mathcal{A}_1 \vert = 89$ ansatz operators. The model is defined in Eq.~\eqref{eq:Schwinger} with parameters $J=w=m=1$.}
	\label{fig:Schwinger}
\end{figure}

\subsection{Verifying a DQS of the Lattice Schwinger Model} \label{subsec:Schwinger}

Simulating the time-evolution of lattice gauge theories on digital quantum devices has recently attracted considerable experimental~\cite{Martinez2016} and theoretical~\cite{PhysRevA.95.023604, PhysRevD.101.074512, Bolens2021} interest. 
Here we numerically illustrate that FHL can be applied to verify the DQS of the lattice Schwinger model as implemented in~\cite{Martinez2016}. 
In particular we demonstrate reconstruction of $\hat{H}_\mathrm{F}(\tau)$ up to first order in $\tau$ in presence of the long-range interactions, that emerge from the elimination of the gauge degrees of freedom (see~\cite{Martinez2016}).

The Hamiltonian of the lattice Schwinger model, acting as the target Hamiltonian of the DQS reads
\begin{equation} \label{eq:Schwinger}
    \hat{H}_{\mathrm{Schwinger}} = \hat{H}_{zz} + \hat{H}_{+-} + \hat{H}_z \,\,,
\end{equation}
where
\begin{align*}
    & \hat{H}_{zz} = \frac{J}{2} \sum_{i=1}^{N-2} \sum_{j=i+1}^{N-1} (N-j) \hat{\sigma}_i^z \hat{\sigma}_j^z \equiv \sum_{i<j} J^{zz}(i,j)\hat{\sigma}_i^z \hat{\sigma}_j^z \,\,, \\
    & \hat{H}_{+-} = w \sum_{j=1}^{N-1} (\hat{\sigma}_j^+ \hat{\sigma}_{j+1}^- + \mathrm{H.c.})\,\,,  \\
    & \hat{H}_z = \frac{m}{2} \sum_{j=1}^N (-1)^j \hat{\sigma}_j^z - \frac{J}{2} \sum_{j=1}^{N-1} (j \, \mathrm{mod} \, 2) \sum_{i=1}^j \hat{\sigma}_i^z\,\,.
\end{align*}
In Fig.~\ref{fig:Schwinger} we show the reconstruction error $\Vert\boldsymbol{c}^{\mathrm{ex}}-\boldsymbol{c}^{\mathrm{rec}}\Vert$ and the corresponding reconstructed $J_\mathrm{rec}^{zz}(i,j)$, with $J=w=m=1$ setting the units of energy. As shown in Fig.~\ref{fig:Schwinger}(a), for small $\tau \approx 0.1$ a zeroth order ansatz yields a low reconstruction error, while for larger $\tau$ Trotter errors become more dominant and require a first order ansatz to achieve a faithful reconstruction. When using a first order ansatz, the structure of the reconstructed $J^{zz}(i,j)$ resembles that of the target Schwinger model up to $\tau \approx 0.6$ as can be seen in Fig.~\ref{fig:Schwinger}(c). Moreover, the reconstructed parameters of the first order terms characterize deviations from the target Hamiltonian in the form of Trotter errors. At $\tau \approx 0.8$ Trotterization breaks down which is reflected in the learned coefficients in Fig.~\ref{fig:Schwinger}(d).

\subsection{Design and Verification of Many-Body Gates: Example of the 1D Cluster Hamiltonian} \label{subsec:ClusterModel}

\begin{figure*} [t]
	\centering  
	\includegraphics[width=1.0\linewidth]{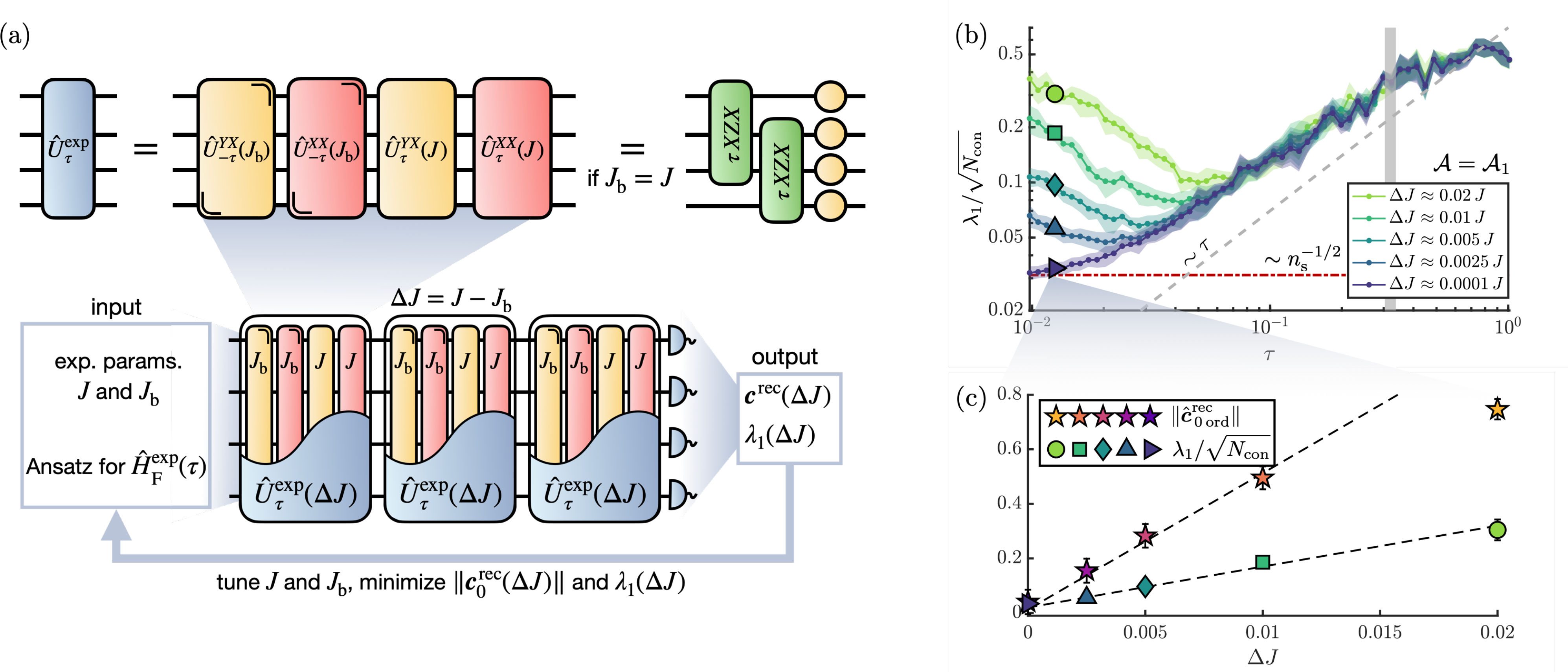}
	\caption{{\bf{Generating 3-body gates}}. (a) Schematic view of FHL applied to the design of multi-body quantum gates, in the specific example of the one-dimensional cluster model studied here, defined in Eq.~\eqref{eq:1DclusterTrotteriz}. (b) Loglog plot of $\lambda_1(\tau)/\sqrt{N_{\mathrm{con}}}$ for different values of the difference $\Delta J$ between forward and backward time-evolution, for an ansatz $\mathcal{A}_1$ containing only the first order terms. Here, $n_{\mathrm{s}}=2000$. As long as $\Delta J>0$, $\lambda_1/\sqrt{N_{\mathrm{con}}}$ cannot show scaling for arbitrarily small values of $\tau$ (as shown by its increase for smaller $\tau$). Thus, $\lambda_1$ can be used as a cost function to be minimized by tuning the experimental parameters controlling $\Delta J$. As shown in (c), minimizing $\lambda_1$ corresponds to minimizing the norm of the zeroth order terms $\Vert\boldsymbol{c}_{\mathrm{rec,0\,ord.}}\Vert$, which can be reconstructed by adding $\mathcal{A}_0$ to the ansatz. Once $\lambda_1/\sqrt{N_{\mathrm{con}}}$ falls below the noise threshold in Eq.~\eqref{eq:Elambda}, we conclude that within our resolution set by $n_{\mathrm{s}}$, $\Delta J\approx 0$. The results are obtained for a system of $N=10$ spins, using $55$ initial product states in random bases, each evolved to $6$ different final times up to a total simulation time of $10/\tau$. The error bars stem from measurement noise and from different initial states realizations.}
	\label{fig:ClusterModel}
\end{figure*}

In this section, we employ our order-by-order learning protocol as a tool to design multi-body interaction gates, i.e., gates coupling more than two qubits, from the resources in \eqref{eq:resources}. Such interactions can be generated from higher order terms in the Magnus expansion, and their strength can be `enhanced' by choosing a Trotterization canceling the lower orders. The cancellation of lower order terms generally requires the ability of reversing the sign of the time-evolution implemented by the resource gates. We consider here the situation in which this inversion can be achieved only approximately, leading to `spurious' low-order terms which we want to correct for. We show that FHL can be integrated in a feedback loop optimization scheme for tuning the experimental control parameters to achieve this correction, thereby enhancing the higher order multi-body target terms.

We numerically exemplify such a protocol in a DQS targeting the one-dimensional cluster Hamiltonian, a prototypical example featuring three-body interactions~\cite{Son2012}. Specifically, our target Hamiltonian reads
\begin{equation}
	\hat{H}_{\mathrm{1DC}}=J_{\mathrm{C}}\sum_{j=1}^{N-2}\hat{\sigma}^x_j\hat{\sigma}^z_{j+1}\hat{\sigma}^x_{j+2}+\sum_{j=1}^{N}B^z_j\hat{\sigma}^z_{j} \,\,.
	\label{eq:1DclusterTarget}
\end{equation}
A Trotter sequence implementing this model with the resources given in \eqref{eq:resources} reads
\begin{equation}
	\hat{U}_{\tau}=\hat{U}^{XX}_{\tau}\,\hat{U}^{YX}_{\tau}\,\hat{U}^{XX}_{-\tau}\,\hat{U}^{YX}_{-\tau} \,\,,
	\label{eq:1DclusterTrotteriz}
\end{equation}
where 
\begin{align*}
    & \hat{U}^{XX}_{\tau}=\hat{U}^{XX}_{\tau}(J)=\mathrm{e}^{-\mathrm{i}\tau J\sum_{j}\hat{\sigma}^x_{j}\hat{\sigma}^x_{j+1}} \,\,,\\
    & \hat{U}^{YX}_{\tau}=\hat{U}^{YX}_{\tau}(J)=\mathrm{e}^{-\mathrm{i}\frac{\tau J}{2}\hat{H}^{\mathrm{odd}}_{YX}}\,\mathrm{e}^{-\mathrm{i}\tau J \hat{H}^{\mathrm{even}}_{YX}}\,\mathrm{e}^{-\mathrm{i}\frac{\tau J}{2}\hat{H}^{\mathrm{odd}}_{YX}} \,\,,
\end{align*}
with $\hat{H}^{\mathrm{odd}}_{YX}=\sum_{j\,\mathrm{odd}}\hat{\sigma}^y_{j}\hat{\sigma}^x_{j+1}$ (similar for $\hat{H}^{\mathrm{even}}_{YX}$). These terms can be generated from the resources \eqref{eq:resources} as $\mathrm{e}^{-\mathrm{i}\tau J\hat{\sigma}^y_{j}\hat{\sigma}^x_{j+1}}=\hat{R}^{(j)}_z(\tfrac{\pi}{4})\,\mathrm{e}^{-\mathrm{i}\tau J\hat{\sigma}^x_{j}\hat{\sigma}^x_{j+1}}\hat{R}^{(j)}_z(\tfrac{-\pi}{4})$. The Floquet Hamiltonian resulting from this Trotterization reads (see Appendix \ref{app:1DCluster} for details)
\begin{equation}
	\hat{H}_{\mathrm{F}}(\tau)=2J\tau\sum_{j=1}^{N-2}\hat{\sigma}^x_j\hat{\sigma}^z_{j+1}\hat{\sigma}^x_{j+2}+2J\tau\sum_{j=1}^{N-1}\hat{\sigma}^z_{j} + \mathcal{O}(\tau^2) \,\,,
	\label{eq:1DclusterFMHam}
\end{equation}
where the zeroth order terms in $\tau$ vanish. We point out that the addition of the $\hat{\sigma}^z$ fields is a consequence of the Trotterization and is not essential: it can be easily canceled by adding $\hat{\sigma}^z$ fields with opposite signs. 

We consider the situation in which the coefficients $J$ of the forward ($\hat{U}^{XX}_{\tau}$ and $\hat{U}^{YX}_{\tau}$) and backward ($\hat{U}^{XX}_{-\tau}$ and $\hat{U}^{YX}_{-\tau}$) evolutions slightly differ from each others, leading to a imperfect cancellation of the zeroth order terms $XX$ and $YX$. We denote with $J_{\rm b}$ the coefficients of the backward evolution operators, and define $\Delta J=|J-J_{\rm b}|$. While here this difference is the same for each gate, the following results also apply to the more general case of non-uniform deviations. As before, the ansatz set to zeroth order is denoted by $\mathcal{A}_0=\{XX,YX\}$, and the ansatz set for the first order terms by $\mathcal{A}_1=\{XZX,Z\}$, i.e., containing the target terms. In order to assess whether the zeroth order terms in $\mathcal{A}_0$ are canceled and $\mathcal{A}_1$ captures the leading terms in $\hat{H}_{\mathrm{F}}(\tau)$, we measure the learning error $\lambda_1(\tau)$ and we show that its value can be used as a \emph{cost function} to be minimized by optimizing the experimental parameters controlling $\Delta J$. The protocol is summarized in Fig.~\ref{fig:ClusterModel}(a), and reads as follows:
\begin{enumerate} 
	\item[(1)] Start from an initial guess for the backward evolution parameters (resulting in a generally non-zero $\Delta J$).
	\item[(2)] Measure $\lambda_1$ as a function of $\tau$ for a first order ansatz $\mathcal{A}_1$.
	\item[(3)] If the scaling is not satisfied down to the noise threshold \eqref{eq:Elambda}, add $\mathcal{A}_0$ to the ansatz and measure the corresponding zeroth order coefficients $\boldsymbol{c}^{\mathrm{rec}}_{\mathrm{0\,ord.}}$.
	\item[(4)] Use the result to optimize the backward evolution parameters in order to minimize $\lambda_1$, and thereby the norm $\Vert\boldsymbol{c}^{\mathrm{rec}}_{\mathrm{0\,ord.}}\Vert$.
	\item[(5)] Repeat from 2.~until scaling down to the noise threshold is observed.
\end{enumerate}
The results of this procedure for the DQS of the one-dimensional cluster model are shown in Fig.~\ref{fig:ClusterModel}(b)-(c) for a system of $10$ qubits, with $J=1$ setting the units of energy. In Fig.~\ref{fig:ClusterModel}(b) we show the behavior of $\lambda_1$ with $\tau$ for different values of $\Delta J$. As expected, for a finite $\Delta J$ scaling cannot be observed down to arbitrarily small values of $\tau$. Instead, as the Trotter step decreases, one observes (i) an initial decrease of $\lambda_1$ for larger values of $\tau$, due to the first order terms in the Floquet Hamiltonian being larger than the (small) spurious zeroth order ones, and (ii) an increase of $\lambda_1$ for smaller values of $\tau$ due to the zeroth order terms becoming dominant. When $\Delta J$ becomes sufficiently small, the effect of the non-canceled zeroth order terms becomes too small to be resolved in the chosen range of Trotter steps. Additional details on this example, in particular the linear scaling of $\lambda_1(\tau)$, can be found in Appendix \ref{app:1DCluster}. In Fig.~\ref{fig:ClusterModel}(c) we show the behavior of $\lambda_1$ as a function of $\Delta J$ for a fixed value of $\tau$, where it is shown to reflect the behavior of the norm of the zeroth order coefficients $\boldsymbol{c}^{\mathrm{rec}}_{\mathrm{0\,ord.}}$. This confirms that the learning error $\lambda_1$ can be used as an experimentally measurable cost function for verifying and optimizing the implementation of target Hamiltonians.

\section{Floquet Liouvillian Learning in Trotterized DQS}
\label{sec:LiouvillianLearning}

In this section, we extend our order-by-order FHL to the verification of DQS subject to dissipative processes, by characterizing the generator of Trotter blocks in terms of a Floquet-Liouvillian. In order to describe this approach, we follow the logic of section~\ref{sec:HLinDQS} by first introducing Liouvillian learning as the underlying routine of Floquet Liouvillian learning (FLL), where coherent as well as incoherent processes are parametrized by an ansatz in Lindblad form. Second, we demonstrate how the quality of the reconstruction can be verified by scaling relations analogous to Eq.~\eqref{eq:ScalingRelation} which we illustrate via numerical examples.

\subsection{Order-by-order FLL}

The situation we have in mind is a DQS that implements a Trotter sequence in analogy to Eq.~\eqref{eq:trotterblock}, where the generator of each gate, instead of being a Hamiltonian $\hat{H}_i$, becomes a time-independent Lindblad superoperator $\mathcal{L}_i$ whose action on a state $\hat{\rho}$ is described by 
\begin{align}  \label{eq:LindbladGate}
\begin{split}
    \mathcal{L}_i\hat{\rho}= & -\mathrm{i}[\hat{H}_i,\hat{\rho}]+ \\
    & +\frac{1}{2}\sum_{k} \gamma_{k} \left( [\hat{L}_k \hat{\rho}, \hat{L}_k^\dagger] + [\hat{L}_k, \hat{\rho} \hat{L}_k^\dagger] \right) \,\,,
\end{split}
\end{align}
with the $\hat{L}_k$ being the jump operators describing the dissipative channels, and $\gamma_k$ the associated rates. The $\hat{H}_i$ and $\hat{L}_k$ are generally few-body operators, meaning that physically implemented Lindbladians $\mathcal{L}_i$ can be described by a polynomial number of parameters.

In presence of dissipation, the role of the evolution operator $\hat{U}_\tau$ is taken by a superoperator
\begin{equation} 
    \mathcal{P}_\tau = \mathrm{e}^{\mathcal{L}_\mathrm{F}(\tau) \tau} = \prod_i \mathrm{e}^{\mathcal{L}_i \tau} \,\,,
    \label{eq:LLFloquetOperator}
\end{equation}
which generates stroboscopic time-evolution of any initial density matrix $\hat{\rho}(0)$ according to $\hat{\rho}(n \tau) = \mathcal{P}_\tau^n \hat{\rho}(0)$. In the following, we describe a method for achieving the order-by-order reconstruction of the generator $\mathcal{L}_\mathrm{F}(\tau)$ of the stroboscopic dynamics.

For sufficiently small $\tau$, $\mathcal{L}_\mathrm{F}(\tau)$ can be expressed as a Magnus expansion in powers of $\tau$
\begin{equation}
    \mathcal{L}_\mathrm{F}(\tau)=\sum_{l=0}^{\infty}\tau^l\mathcal{Q}_l \,\,,
    \label{eq:LLMagnusExp}
\end{equation}
with $\mathcal{Q}_0=\sum_i\mathcal{L}_i$ and $\mathcal{Q}_{l}$ ($l>0$) given by $l$-nested commutators of the $\mathcal{L}_i$ in complete analogy with the unitary case. 
As a consequence, a finite truncation of the above expansion for $\mathcal{L}_\mathrm{F}(\tau)$ is still a local operator, and can be learned efficiently as we explain below. For more details on the Floquet Liouvillian, including an explicit formula for the commutator of any two Lindblad superoperators, see Ref.~\cite{Schnell2021}.

As in the FHL case, the starting point of the FLL protocol is an ansatz for the experimentally realized $\mathcal{L}^\mathrm{exp}_\mathrm{F}(\tau)$, which can be chosen of the form~\cite{Schnell2021}
\begin{align} \label{eq:LLansatz}
    \mathcal{L}_{\mathcal{A}}&(\boldsymbol{c}^H,\boldsymbol{c}^\mathcal{D})\hat{\rho} = -\mathrm{i} \sum_j c_j^H [\hat{h}_j,\hat{\rho}] \nonumber\\
    &+\frac{1}{2} \sum_{l,m} c_{l,m}^\mathcal{D} \left( [\hat{\ell}_l \hat{\rho}, \hat{\ell}_m^\dagger] + [\hat{\ell}_l, \hat{\rho}\,\hat{\ell}_m^\dagger] \right) \,\,,
\end{align}
specified by a set of operators $\mathcal{A}=\mathcal{A}_H\cup\mathcal{A}_{\mathcal{D}}=\{\hat{h}_j\}_j\cup\{\hat{\ell}_k\}_k$ and depending on parameters $\boldsymbol{c}^H$ and $\boldsymbol{c}^\mathcal{D}$ to be determined. To generate constraints for reconstructing the coefficients $\boldsymbol{c}^H$ and $\boldsymbol{c}^\mathcal{D}$ we impose the validity of the Ehrenfest theorem, in the spirit of Refs.~\cite{Bairey2019,Bairey2020}. Specifically, for any observable $\hat{A}$, initial state $\hat{\rho}(0)$ and final time $n\tau$, it holds that
\begin{equation} \label{eq:LLcond}
    \langle\hat{A}\rangle_{n\tau} - \langle\hat{A}\rangle_0 = \int_0^{n\tau} \tr(\hat{A}\, \mathcal{L}^\mathrm{exp}_\mathrm{F}(\tau) \hat{\rho}(t)) \, \mathrm{d}t \,\,,
\end{equation}
with $\langle\cdot\rangle_t=\tr(\cdot\,\hat{\rho}(t))$, where the above integral is necessarily discretized since the integrand is measured only at stroboscopic times. Thus, the Liouvillian parameters can be determined by choosing $N_{\rm con}>N_{\mathcal{A}}$ constraint operators $\hat{A}_k$ and by minimizing the sum of the residuals
\begin{align}
\begin{split}
    \Delta_k(\boldsymbol{c}^H,\boldsymbol{c}^\mathcal{D})=&\,\int_0^{n\tau} \tr(\hat{A}_k\,\mathcal{L}_{\mathcal{A}}(\boldsymbol{c}^H,\boldsymbol{c}^\mathcal{D})\hat{\rho}(t)) \, \mathrm{d}t \\
    &-\langle\hat{A}_k\rangle_{n\tau} + \langle\hat{A}_k\rangle_0 \,\,.
\end{split}
\end{align}
This is a linear optimization problem which can be recast in matrix form as
\begin{equation} \label{eq:LLcostfct}
    (\boldsymbol{c}^H_{\mathrm{rec}},\boldsymbol{c}^{\mathcal{D}}_{\mathrm{rec}}) = \mathrm{arg}\min_{\boldsymbol{c}^H,\boldsymbol{c}^{\mathcal{D}}} \Vert G^H \boldsymbol{c}^H + G^{\mathcal{D}} \boldsymbol{c}^{\mathcal{D}} - \boldsymbol{b} \Vert \,\,,
\end{equation}
where the matrices $G^H$ and $G^{\mathcal{D}}$ and the vector $\boldsymbol{b}$ are determined from measured expectation values as
\begin{align}
    &G^H_{k,j} = \int_0^{n\tau} \langle-\mathrm{i} [\hat{A}_k,\hat{h}_j]\rangle_t\,\mathrm{d}t \label{eq:GintH}\,\,,\\
    &G^{\mathcal{D}}_{k,(l,m)} = \frac{1}{2}\int_0^{n\tau} \langle[\hat{\ell}_l^\dagger,\hat{A}_k]\hat{\ell}_m\rangle_t + \langle\hat{\ell}_l^\dagger[\hat{A}_k,\hat{\ell}_m]\rangle_t\,\mathrm{d}t \label{eq:GintD}\,\,,\\
    &b_k = \langle\hat{A}_k\rangle_{n\tau} - \langle\hat{A}_k\rangle_{0} \label{eq:bconstr}\,\,.
\end{align}

The role of $\lambda_1$ as learning error is taken by the minimum of the cost function in Eq.~\eqref{eq:LLcostfct}, which we denote by
\begin{equation}
    \Delta = \Vert G^H \boldsymbol{c}_{\rm rec}^H + G^{\mathcal{D}} \boldsymbol{c}_{\rm rec}^{\mathcal{D}} - \boldsymbol{b} \Vert \,\,.
\end{equation}
As for FHL in Sec.~\ref{sec:HLinDQS}, we study $\Delta$ as a function of the Trotter step for a fixed ansatz. We obtain a scaling relation analogous to Eq.~\eqref{eq:ScalingRelation} for the reconstructed Liouvillian $\mathcal{L}_{\mathrm{rec}}(\tau)=\mathcal{L}_{\mathcal{A}}(\boldsymbol{c}^H_\mathrm{rec}(\tau),\boldsymbol{c}^{\mathcal{D}}_\mathrm{rec}(\tau))$, namely
\begin{equation} \label{eq:LLScalingRelation}
 \Delta(\tau) \sim \tau^L \Leftrightarrow \Vert \mathcal{L}_{\mathrm{rec}}(\tau) - \mathcal{L}^\mathrm{exp}_\mathrm{F}(\tau)  \Vert \sim \tau^L \,\,,
\end{equation}
which we prove in App.~\ref{app:ScalingLiouvillian}. In analogy to order-by-order FHL in Sec.~\ref{sec:HLinDQS}, Eq.~\eqref{eq:LLScalingRelation} can be used to assess whether the chosen ansatz encompasses the experimental Floquet Liouvillian.

\subsection{Numerical Illustration of FLL}

\begin{figure} 
	\centering  
	\includegraphics[width=0.85\linewidth]{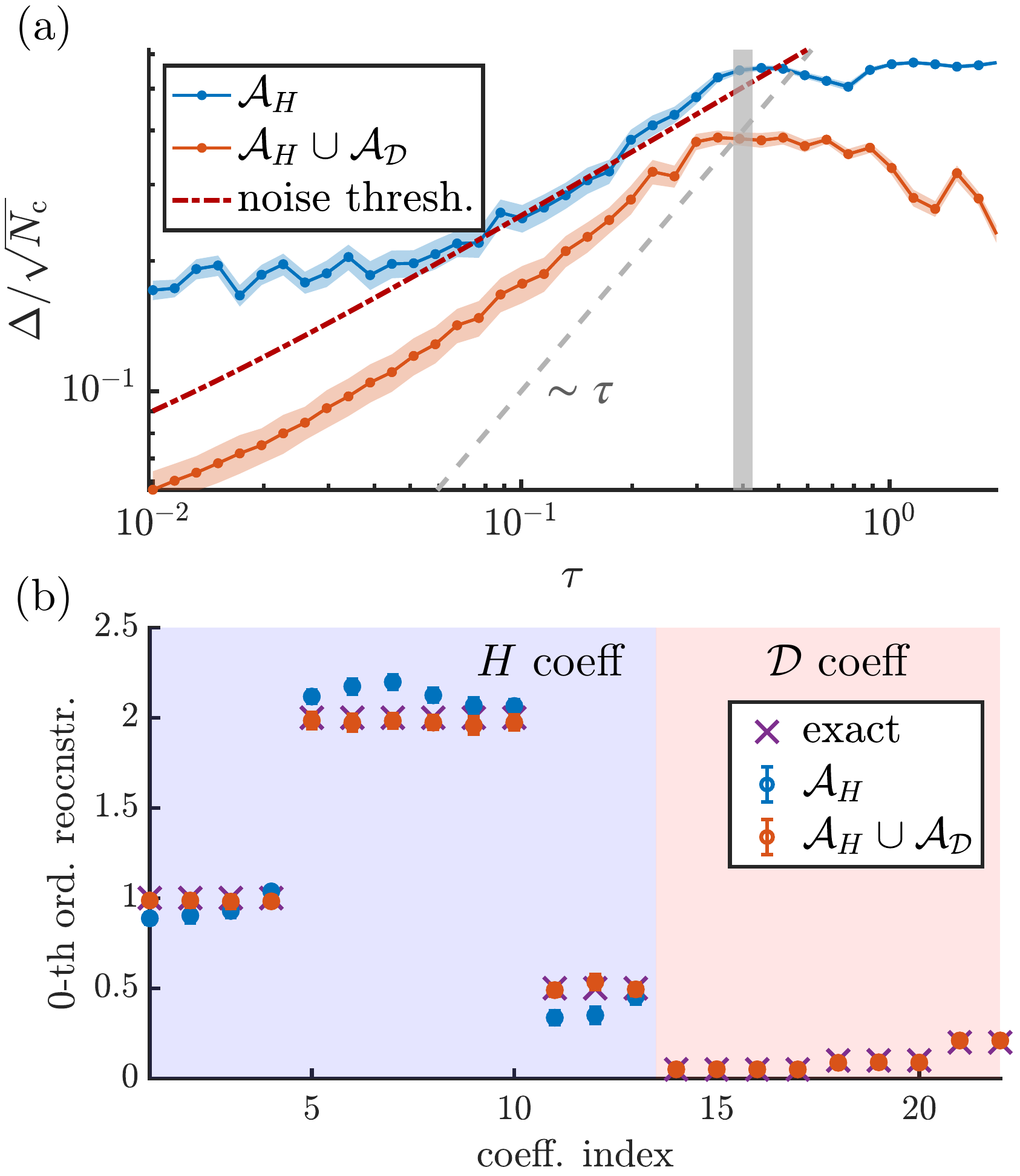}
	\caption{{\bf{Floquet Liouvillian Learning (FLL)}}. FLL in a dissipative DQS of the Heisenberg/XXZ model for $N=4$ spins. (a) $\Delta(\tau)$ for a zeroth order ansatz without (blue) and with (red) dissipative terms. The errorbars stem from a finite measurement budget per operator of $n_{\rm s} = 1000$. The dashed red line gives an upper bound to $\Delta(\tau)$ from measurement noise. The dashed grey line is a guide to the eye that shows scaling linear in $\tau$. (b) Reconstructed parameters for both cases (blue and red dots) together with the exact values (purple crosses). Hamiltonian parameters are defined in Eq.~\eqref{eq:HeisenbergTarget}: $B_x=1$ (indices $1\text{-}4$), $J_{xx}=J_{yy}=2$ (inds.~$5\text{-}10$), $J_z=0.5$ (inds.~$11\text{-}13$). Dissipation rates [Eq.~\eqref{eq:JumpOpsExample}] are $\gamma^{(-)}_j = 0.0125$ (inds.~$14\text{-}17$), $\gamma^{(zz)}_j = 0.025$ (inds.~$18\text{-}20$), $\gamma^{(xxx)}_j=0.05$ (inds.~$21\text{-}22$). The learned dissipation rates are $4$ times their values in $\mathcal{L}_i$, because dissipation acts during the application of all gates, which are four in the present example [see Eq.~\eqref{eq:HeisenbergTrotteriz}]. The constraint operators are chosen to be all single spin Pauli operators $\hat{\sigma}_i^x, \hat{\sigma}_i^y, \hat{\sigma}_i^z$ for a number of $20$ random initial product states each. The discretization of the integral in Eq.~\eqref{eq:LLcond} is chosen to be $\tau$ up to a fixed total simulation time $T=6$.}
	\label{fig:LLofTau}
\end{figure}

In this section we present a numerical illustration of FLL applied to the DQS of the Heisenberg model as given in Eq.~\eqref{eq:HeisenbergTrotteriz}, in presence of dissipation. Specifically, we replace the generators of each individual gate $\hat{U}^i_{\tau}$, with $i=\{X,XX,YY,ZZ\}$ in Eq.~\eqref{eq:HeisenbergTrotteriz}, by a Lindblad superoperator $\mathcal{L}_i$, defined as in Eq.~\eqref{eq:LindbladGate} with jump operators
\begin{equation}
    \hat{L}_k \in \{ \hat{\sigma}_j^-, \hat{\sigma}_j^z \hat{\sigma}_{j+1}^z, \hat{\sigma}_{j-1}^x \hat{\sigma}_j^x \hat{\sigma}_{j+1}^x\} \,\,, 
    \label{eq:JumpOpsExample}
\end{equation}
and corresponding dissipation rates $\gamma_j^{(-)}, \gamma_j^{(zz)}, \gamma_j^{(xxx)}$ for each spin $j$, with Eq.~\eqref{eq:LLFloquetOperator} being the associated evolution superoperator over one Trotter cycle. The dissipative processes in this example are chosen in order to illustrate the capability of FLL to learn multi-qubit jump operators. The characterization of such dissipative processes might give insight into the underlying physical processes and help improve the quantum device. (See for example Ref.~\cite{cosmicrays} on how cosmic radiation affects the coherence time of superconducting qubits.)

Our numerical results are presented in Fig.~\ref{fig:LLofTau}. The blue and red data in Fig.~\ref{fig:LLofTau}(a) show the behavior of $\Delta(\tau)$ for an ansatz missing and including the dissipative terms, 
respectively. In complete analogy to Sec.~\ref{subsec:AdaptiveErrors}, the missing ansatz terms result in a plateau of $\Delta(\tau)$ for small $\tau$. Only when the dissipative terms are included in the ansatz the scaling is observed down to small values of $\tau$, hence verifying the complete reconstruction to zeroth order. The red dashed line in Fig.~\ref{fig:LLofTau}(a) corresponds to the noise threshold, providing an upper bound on $\Delta(\tau)$ in presence of measurement noise (see Appendix \ref{app:Elambda} for a derivation). The blue data being larger than this bound implies the presence of terms not accounted for in the corresponding ansatz. In Fig.~\ref{fig:LLofTau}(b) we plot the reconstructed parameters for both Ansätze (blue and red dots), together with the exact parameters (purple crosses). Note that adding dissipative terms to the ansatz not only allows for reconstructing the corresponding coefficients, but also improves the accuracy of the learned Hamiltonian parameters.

\section{Summary and Outlook} \label{subsec:Conclusion}

In this work we proposed an efficient method for process tomography of Trotter blocks in DQS as an experimental protocol for the characterization and verification of NISQ devices. The protocol is based on `learning' either the Floquet Hamiltonian $\hat H_{\rm F}^\mathrm{exp}(\tau)$ as generator of unitary evolution for a Trotter step $\tau$, or the Floquet Liouvillian ${\cal L}_{\rm F}^\mathrm{exp}(\tau)$ in the case of dissipative dynamics.
The Floquet Hamiltonian, and similarly the Liouvillian, admits an expansion in powers of the Trotter step as $\hat{H}_{\mathrm{F}}^{{\rm exp}}(\tau)=\sum_{l=0}^{\infty}\hat{\Omega}_{l}^{{\rm exp}}\,\tau^{l}$.
Owing to this structure, our learning protocol starts from an ansatz for the operator content of a truncation of $\hat H_{\rm F}^\mathrm{exp}(\tau)$ or ${\cal L}_{\rm F}^\mathrm{exp}(\tau)$ to a chosen order $\tau^L$.
The ansatz contains theoretically expected operators (e.g., the target Hamiltonian of the DQS) together with terms associated with implementation errors or dissipative channels that one wants to test for.
The corresponding coefficients are reconstructed by fitting the ansatz to measurement data from a set of time-evolved states, for several values of $\tau$. This procedure comes with an associated, measurable learning error $\lambda_1(\tau)$, whose behavior as a function of $\tau$ (i) reveals up to which order in $\tau$ the chosen ansatz reconstructs the implemented Floquet $\hat{H}_\mathrm{F}^\mathrm{exp}(\tau)$ or ${\cal L}_{\rm F}^\mathrm{exp}(\tau)$, (ii) provides means of systematically improving the ansatz if needed, and (iii) enables the detection of the Trotter threshold $\tau_{*}$, i.e., the point at which the Trotter approximation breaks down and $\hat{H}_\mathrm{F}^\mathrm{exp}(\tau)$ cannot be described by a power-expansion in $\tau$.
This protocol is scalable, with a number of required measurement runs scaling polynomially with system size (see also Appendix \ref{app:scalability}).

With our procedure, Trotter errors, control errors as well as dissipative processes affecting an experimental DQS can be learned and distinguished from one another. These direct insights into the microscopic processes underlying the implemented dynamics offer a potential route for compensation of errors via fine tuning, e.g., via a feedback loop.

Furthermore, the present techniques provide ingredients for the design of complex quantum gates or many-body interactions (see Sec.~\ref{subsec:ClusterModel}) and subsequent verification of the elementary Trotter blocks.
A natural extension of our approach for tailoring $N$-body interactions in DQS is the integration of the protocol into a variational feedback-loop scheme for quantum gate design. In contrast to traditional approaches to \emph{quantum compiling}~\cite{Khatri2019,Moro2021,Bolens2021,Volkoff2021,Jones2022}, which optimize the fidelity with respect to some target unitary, here one exploits Hamiltonian learning to directly minimize the Hamiltonian distance to design a desired target Hamiltonian. 

We note that the sample complexity of our protocol, i.e., the number of required experimental runs, can be reduced significantly with efficient parametrization schemes of Hamiltonians or Liouvillians. In particular, one can devise adaptive schemes similar to the one of Sec.~\ref{subsec:AdaptiveErrors}, where the ansatz parametrization is improved \emph{on the fly} as new measurements are collected. Other promising strategies for reducing the sample complexity involve the preparation of (optimized) entangled states as inputs to the Hamiltonian or Liouvillian learning routine (see for instance Ref.~\cite{Dutt2021}). 

Finally, we emphasize that while the examples considered in the present work have focused on DQS of quantum many-body models motivated by condensed matter physics, the technique finds immediate application to the simulation of quantum chemistry problems~\cite{whitfield2011}, or the real time dynamics of lattice gauge theories~\cite{Martinez2016,PhysRevD.100.034518, PhysRevA.95.023604, GonzalezCuadra2022}.

\section*{Acknowledgements}
We thank A.~Daley, R.~van Bijnen, L.~M.~Sieberer and L.~K.~Joshi for valuable discussions. 
Research  is supported by the US Air Force Office of Scientific Research (AFOSR) via IOE Grant No.~FA9550-19-1-7044 LASCEM, the European Union’s Horizon 2020 research and innovation program under Grant Agreement No.~817482 (PASQuanS), and by the Simons Collaboration on Ultra-Quantum Matter, which is a grant from the Simons Foundation (651440, P.Z.), and by the Institut f\"ur Quanteninformation. Innsbruck theory is a member of the NSF Quantum Leap Challenge Institute Q-Sense. The computational results presented have been achieved (in part) using the HPC infrastructure LEO of the University of Innsbruck.

\appendix

\section{The Magnus expansion}
\label{app:Magnus}

In this appendix, we write down the first few orders of the Magnus expansion Eq.~\eqref{eq:hf}, and comment on the structure of $\hat H_{\rm F}(\tau)$ for sufficiently small $\tau< \tau_\ast$ and its relevance for Floquet Hamiltonian Learning. For the Trotter decomposition in Eq.~\eqref{eq:trotterblock}, the operators $\hat{\Omega}_{l}$ in Eq.~\eqref{eq:hf} consist of $l$-nested commutators
of the Hamiltonian components $\hat{H}_{j}$
according to the Baker-Campbell-Hausdorff
formula~\cite{Casas2009},
\begin{align}
\hat{\Omega}_{0} & =\sum_{j}\hat{H}_{j}\,\,,\label{eq:FMorders0}\\
\hat{\Omega}_{1} & =-\frac{\mathrm{i}}{2}\sum_{i<j}[\hat{H}_{j},\hat{H}_{i}]\,\,, \label{eq:FMorders1}\\
\begin{split}\hat{\Omega}_{2} & =\frac{1}{12}\sum_{i<j}\Big[\sum_{l<j}\hat{H}_{l}-\hat{H}_{j},[\hat{H}_{j},\hat{H}_{i}]\Big]\\
 & -\frac{1}{4}\sum_{i<l<j}[\hat{H}_{j},[\hat{H}_{l},\hat{H}_{i}]]\,\,,\ \text{etc.}
\end{split}
 \label{eq:FMorders2}
\end{align}
Here, $\hat{\Omega}_0$ is the target Hamiltonian Eq.~\eqref{eq:h}, which we assumed to be local (see main text).
Then, due to the nested commutator structure in Eqs.~(\ref{eq:FMorders0})-(\ref{eq:FMorders2}) (and higher orders), the $\hat{\Omega}_l$ for $l>0$ are still local with an increasing spatial support. Therefore, in the context of Hamiltonian learning, for $\tau<\tau_\ast$, there exists an ansatz for a finite truncation of $\hat{H}_\mathrm{F}(\tau)$ with number of parameters that scales polynomially in system size.


This also holds in a realistic setting, where the implemented Floquet Hamiltonian $\hat{H}_\mathrm{F}^\mathrm{exp}(\tau)$ will be affected by experimental imperfections.
In this case the higher orders of the Magnus expansion cannot be calculated from the theoretical Trotter gate sequence, as in the ideal case, since the location and type of errors is unknown. Nevertheless, the nested commutators in Eqs.~(\ref{eq:FMorders0})-(\ref{eq:FMorders2}) \emph{etc.} can provide intuition for the structure of an ansatz for $\hat{H}_\mathrm{F}^\mathrm{exp}(\tau)$ up to a given order in $\tau$, that is then extended by the expected error terms (and nested commutators thereof) as we do in the main text.

\section{Reconstructing the overall scale of the Hamiltonian}
\label{app:OverallScale}

\begin{figure} 
	\centering  
	\includegraphics[width=\linewidth]{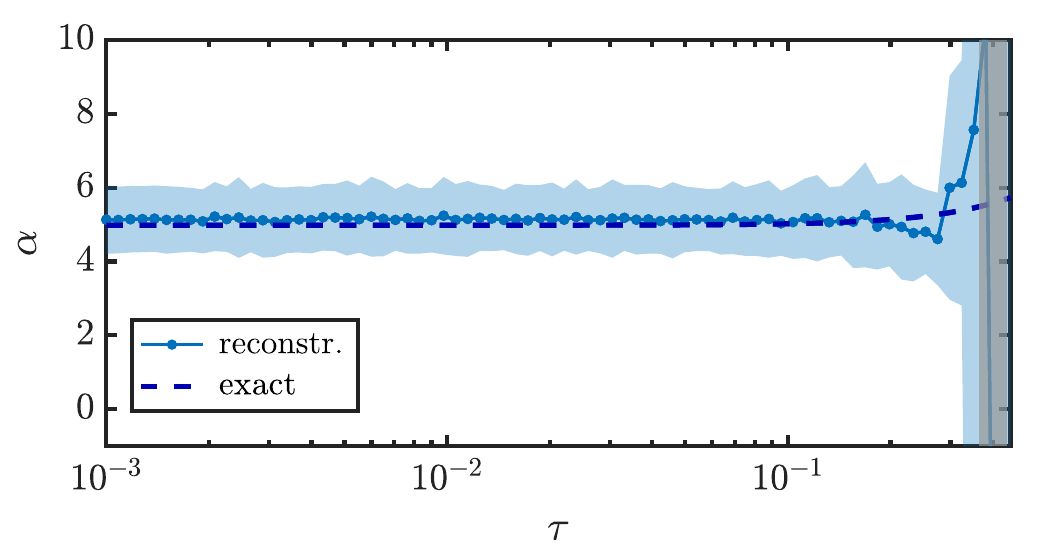}
	\caption{{\bf{Overall scale reconstruction}}. Reconstruction of overall scale for the first order learning shown in Fig.~\ref{fig:Heisenberg}. The initial state chosen here is a N\`eel state evolved up to a total time of $n\tau=5$, with constraint operator $\hat{A}=\hat{\sigma}^z_1$. The error bars stem from simulated measurement noise with $n_{\mathrm{s}}=300$.}
	\label{fig:OverallScale}
\end{figure}

In this appendix we explain how to determine the overall scale of a Hamiltonian that has been reconstructed using FHL as described in Sec.~\ref{sec:HLinDQS}. As discussed in the main text, the protocol of Ref.~\cite{Li2020} can determine the Hamiltonian parameters only up to an undetermined multiplicative factor $\alpha(\tau)$, since $\boldsymbol{c}^{\mathrm{rec}}(\tau)$ and $\alpha(\tau)\,\boldsymbol{c}^{\mathrm{rec}}(\tau)$ are equivalent solutions of the minimization problem \eqref{eq:argmin}. The overall-scale reconstruction derives from the HL protocol proposed in Ref.~\cite{Bairey2019}, which determines the Hamiltonian from measurements of time-resolved local observables. Likewise, here we determine $\alpha(\tau)$ based on the fact that, for any observable $\hat{A}$ it holds that
\begin{equation}
	\langle \hat{A} \rangle_{n\tau} - \langle \hat{A} \rangle_0 = \int_0^{n\tau} \langle -\mathrm{i} [\hat{A},  \hat{H}^\mathrm{exp}_{\mathrm{F}}(\tau)] \rangle_t \; \mathrm{d} t \,\,,
	\label{eq:IntegralEhrenfestObservable}
\end{equation}
for any time $n\tau$, where $\langle\cdot\rangle_{n\tau}=\langle\psi(n\tau)|\cdot|\psi(n\tau)\rangle$ with $|\psi(n\tau)\rangle=(\hat{U}^\mathrm{exp}_{\tau})^n|\psi(0)\rangle$ for any chosen $|\psi(0)\rangle$. In the above equation, there is an inherent discretization error associated with the integral, which comes from the fact that expectation values are measured at integer multiples of $\tau$. These errors can be suppressed by the choice of the integration routine, and become negligible compared to measurement noise and Trotter errors for small $\tau$. For a reconstruction up to order $L$ the Floquet Hamiltonian $\hat{H}^\mathrm{exp}_{\mathrm{F}}(\tau)$ fulfills
\begin{equation}
	\hat{H}^\mathrm{exp}_{\mathrm{F}}(\tau) = \alpha(\tau) \sum_j c^{\mathrm{rec}}_j(\tau) \hat{h}_j + \mathcal{O}(\tau^{L+1}) \,\,,
\end{equation}
with $c^{\mathrm{rec}}_j(\tau)$ being the parameters learned from the protocol in Sec.~\ref{sec:HLinDQS}, and $\alpha(\tau)$ the prefactor we want to reconstruct. Using Eq.~\eqref{eq:IntegralEhrenfestObservable} we can obtain $\alpha$ as follows
\begin{equation}
	\alpha(\tau) = \frac{ \langle \hat{A} \rangle_{n\tau} - \langle \hat{A} \rangle_0 } {\sum_j c^{\rm rec}_j(\tau) \int_0^{n\tau} \langle -\mathrm{i} [\hat{A},  \hat{h}_j] \rangle_t \; \mathrm{d} t} + \mathcal{O}(\tau^{L+1}) \,\,.
	\label{eq:OverallScaleRec}
\end{equation}
We illustrate an application of this method in Fig.~\ref{fig:OverallScale}, where we plot the reconstructed $\alpha(\tau)$ as a function of $\tau$ for the first order learning shown in Fig.~\ref{fig:Heisenberg}. In this example we used a fourth order integration routine. As a result, the discretization errors in evaluating the integral in Eq.~\eqref{eq:OverallScaleRec} are negligible for all values of $\tau$ up to approximately the Trotterization threshold. Thus, with this method we can accurately reconstruct $\alpha$ for each $\tau<\tau_{*}$ (the shaded vertical region in the figure). We expect this behavior to be rather generic: denoting with $J$ the dominant energy scale in $\hat{H}^\mathrm{exp}_{\rm F}(\tau)$, in general we would have $\tau_{*}\leq J^{-1}$. Since $J^{-1}$ is the fastest timescale at which the expectation values $\langle -\mathrm{i}[\hat{A},\hat{h}_j] \rangle_t$ change, we expect the $\tau$ discretization for $\tau<\tau_{*}\leq J^{-1}$, together with a high-order integration routine, to be sufficient to suppress integration errors.

\section{Proof of the scaling relation in Floquet Hamiltonian Learning} \label{app:ScalingRelation}

In this appendix we prove the scaling relation Eq.~\eqref{eq:ScalingRelation}, that forms the basis of our order-by-order learning procedure, in the context of unitary dynamics. Its extension to dissipative DQS is proven in App.~\ref{app:ScalingLiouvillian}.

Consider a DQS that implements a Trotter block $\hat{U}^\mathrm{exp}_\tau$. Based on an ansatz $\mathcal{A}$ and a set of initial states $\{ \vert \psi_m \rangle \}_m$ one obtains a constraint matrix $M_\mathcal{A}$ and reconstructs $\hat{H}_{\mathrm{rec}}(\tau)$. In the following, we prove the following scaling relation
\begin{equation} \label{aeq:ScalingRelation}
	\lambda_1(\tau) \in \mathcal{O}(\tau^{K+1}) \Leftrightarrow  \left[ \hat{H}_\mathrm{rec}(\tau) , \hat{U}^\mathrm{exp}_\tau \right]  \in \mathcal{O}(\tau^{K+1}) \,\,,
\end{equation}
where $\lambda_1$ is the lowest singular value of $M_\mathcal{A}$. Here $\mathcal{O}(\tau^k) = \{ f(\tau) = \sum_{l=k}^\infty a_l \tau^l \vert a_l \in \mathbb{C} \}$. Note that for any functions $f(\tau),g(\tau)$ it holds that
\begin{equation*}
\begin{split}
	[f(\tau) \in \mathcal{O}(\tau^k) \Leftrightarrow g(\tau) \in \mathcal{O}(\tau^k)] \Leftrightarrow \\
	\Leftrightarrow [f(\tau) \sim \tau^k \Leftrightarrow g(\tau) \sim \tau^k] \,\,.
	\end{split}
\end{equation*}
At the end of this section we comment on how Eq.~\eqref{aeq:ScalingRelation} relates to Eq.~\eqref{eq:ScalingRelation} in the main text. During the proof we will drop the superscript `$\mathrm{exp}$' in $\hat{U}^\mathrm{exp}_\tau$ to improve readability of the equations.

To prove Eq.~\eqref{aeq:ScalingRelation} we first note that 
\begin{equation}
	\lambda_1 \sim \tau^k  \Leftrightarrow M_\mathcal{A} \boldsymbol{c} \sim \tau^k \,\,, 
	\label{aeq:lambdascaling}
\end{equation}
because $\lambda_1 = \Vert M_\mathcal{A} \boldsymbol{c} \Vert$. Moreover we find that
\begin{equation}
		(M_\mathcal{A} \boldsymbol{c})_m = \langle \psi_m  \vert \hat{H}_\mathrm{rec} - (\hat{U}_\tau^n)^\dagger \hat{H}_\mathrm{rec} \hat{U}_\tau^n \vert \psi_m \rangle \,\,,
		\label{aeq:Mc_expr}
\end{equation}
where $\vert \psi_m \rangle$ the initial state that defines the constraint. For simplicity we set $n=1$ in the following.

We proceed by showing that for any operator $\hat{X}$ one has 
\begin{equation} \label{aeq:ScalRel}
\begin{split}
	\forall m : \, \langle \psi_m  \vert \hat{X} - \hat{U}_\tau^\dagger \hat{X} \hat{U}_\tau \vert \psi_m \rangle \in \mathcal{O}(\tau^{k}) \Leftrightarrow \\
	\Leftrightarrow [\hat{X},\hat{U}_\tau] \in \mathcal{O}(\tau^{k}) \,\,,
	\end{split}
\end{equation}
which, together with Eq.~\eqref{aeq:lambdascaling} and \eqref{aeq:Mc_expr}, will prove the scaling relation \eqref{aeq:ScalingRelation} for the commutator. We do this in two steps.\\
\,\\
\textbf{Step 1}: \emph{Given an operator $X$ we show that $\hat{X} - \hat{U}_\tau^\dagger \hat{X} \hat{U}_\tau \in \mathcal{O}(\tau^k) \Leftrightarrow [\hat{X}, \hat{U}_\tau ] \in \mathcal{O}(\tau^k)$.} 

The unitary $\hat{U}_\tau$ is analytic in $\tau$ for all $\tau$ and hence each element can be written $(\hat{U}_\tau)_{i,j} = \sum_l u_{i,j,l} \tau^l$ for $u_{i,j,l} \in \mathbb{C}$. Because $\hat{U}_\tau$ acts linearly it follows that $\hat{A} \in \mathcal{O}(\tau^k) \Leftrightarrow \hat{U}_\tau \hat{A} \in \mathcal{O}(\tau^k) \Leftrightarrow  \hat{A} \hat{U}_\tau \in \mathcal{O}(\tau^k)$ for any operator $\hat{A}$. The same is true for $\hat{U}_\tau^\dagger$. This proves the statement since $\hat{U}_\tau^\dagger [\hat{U}_\tau,\hat{X}] = \hat{X} - \hat{U}_\tau^\dagger \hat{X} \hat{U}_\tau$.\\
\,\\
\textbf{Step 2}: \emph{Given an operator $X$ we show that  $\hat{X} \in \mathcal{O}(\tau^k) \Leftrightarrow \forall m: \, \langle \psi_m \vert \hat{X} \vert \psi_m \rangle \in \mathcal{O}(\tau^k)$.}

Direction ``$\Rightarrow$" follows because taking the expectation value is a linear operation that cannot introduce lower orders in $\tau$.

To prove direction ``$\Leftarrow$" we split $\hat{X}=\hat{X}_0+\hat{X}_\tau$ with $\hat{X}_0 = \sum_{j=0}^{k-1} \hat{x}_j \tau^j$ a polynomial in $\tau$ and $\hat{X}_\tau \in \mathcal{O}(\tau^k)$. From the assumed scaling $\forall m: \, \langle \psi_m \vert \hat{X} \vert \psi_m \rangle \in \mathcal{O}(\tau^k)$ it follows $\forall m: \, \langle \psi_m \vert \hat{X}_0 \vert \psi_m \rangle \in \mathcal{O}(\tau^k)$. Moreover, because taking the expectation value is a linear operation, it cannot increase the scaling beyond the maximal power in $\hat{X}_0$, which is $\tau^{k-1}$ and hence $\forall m: \, \langle \psi_m \vert \hat{X}_0 \vert \psi_m \rangle = 0$. Then if $\hat{X}_0 \neq 0$ we can choose an eigenstate $ \hat{X}_0 \vert u \rangle = u \vert u \rangle$ with $u \neq 0$. This state fulfills $\langle u \vert \hat{X}_0 \vert u \rangle = u \neq 0$ and choosing sufficient constraints such that $\exists\,\tilde{m}: \, \vert u \rangle = \vert \psi_{\tilde{m}} \rangle$ proves that $\hat{X}_0=0$ via contradiction. This means that $\hat{X} = \hat{X}_\tau \in \mathcal{O}(\tau^k)$. Note that the proof assumes that $\vert u \rangle$ is not an eigenstate of $\hat{U}_\tau$, since such a state would result in a row of zeros in the constraint matrix, and hence does not constitute a valid constraint. We note here that the choice of optimal constraints is a difficult task in general. However, in all the many-body examples presented in the main text, product states, obviously different from eigenstates of $\hat{U}_{\tau}$, were always a good choice as constraints. \\

In the main text we refer to the scaling relation Eq.~\eqref{aeq:ScalingRelation} in the context of the operator distance $\hat{H}_\mathrm{rec}(\tau) - \hat{H}^\mathrm{exp}_\mathrm{F}(\tau)$ between the reconstructed Hamiltonian $\hat{H}_\mathrm{rec}(\tau)$ and the experimental Floquet Hamiltonian $\hat{H}^\mathrm{exp}_\mathrm{F}(\tau)$ (see Eq.~\eqref{eq:ScalingRelation}). This simplification applies in the case, where the ansatz does not contain any conserved quantity of any truncation of the Magnus expansion of $\hat{H}_\mathrm{F}^\mathrm{exp}(\tau)$ (except the truncation of $\hat{H}^\mathrm{exp}_\mathrm{F}(\tau)$ itself). (Note that if these conserved quantities are known, they can be `projected out' from the ansatz.) In that case, using $[\hat{H}_\mathrm{rec}(\tau) ,\hat{U}^\mathrm{exp}_\tau] = [\hat{H}_\mathrm{rec}(\tau) - \hat{H}_\mathrm{F}^\mathrm{exp}(\tau) ,\hat{U}_\tau^\mathrm{exp}]$, it holds that
\begin{equation} 
    \label{aeq:SimpScalRel}
    \begin{split}
	[\hat{H}_\mathrm{rec}(\tau) - \hat{H}^\mathrm{exp}_\mathrm{F}(\tau) ,\hat{U}^\mathrm{exp}_\tau] \in \mathcal{O}(\tau^k) \Leftrightarrow \\
	\Leftrightarrow \hat{H}_\mathrm{rec}(\tau) - \hat{H}^\mathrm{exp}_\mathrm{F}(\tau) \in \mathcal{O}(\tau^k) \,\,.
	\end{split}
\end{equation}
Direction ``$\Leftarrow$" in Eq.~\eqref{aeq:SimpScalRel} follows because the commutation with $\hat{U}^\mathrm{exp}_\tau$, being a linear operation, cannot introduce terms with lower order in $\tau$. To prove direction ``$\Rightarrow$" in Eq.~\eqref{aeq:SimpScalRel}, define $\hat{H}_\mathrm{rec}(\tau) - \hat{H}^\mathrm{exp}_\mathrm{F}(\tau) = \hat{A}_0 + \hat{A}_\tau$ with $\hat{A}_0 = \sum_{l=0}^{k-1} \hat{a}_l \tau^l$ and $\hat{A}_\tau \in \mathcal{O}(\tau^k)$. Then from $[\hat{A}_0 + \hat{A}_\tau ,\hat{U}^\mathrm{exp}_\tau] \in \mathcal{O}(\tau^k)$ it follows that $[\hat{A}_0,\hat{U}^\mathrm{exp}_\tau] \in \mathcal{O}(\tau^k)$ meaning that $\hat{A}_0$ is a conserved quantity of $\hat{U}^\mathrm{exp}_\tau$ to order $k$, different from any truncation of $\hat{H}^\mathrm{exp}_\mathrm{F}(\tau)$.

\section{Trotterization threshold and transition to quantum chaos} \label{app:TrotterThreshold}

\begin{figure}
	\centering  
	\includegraphics[width=0.9\linewidth]{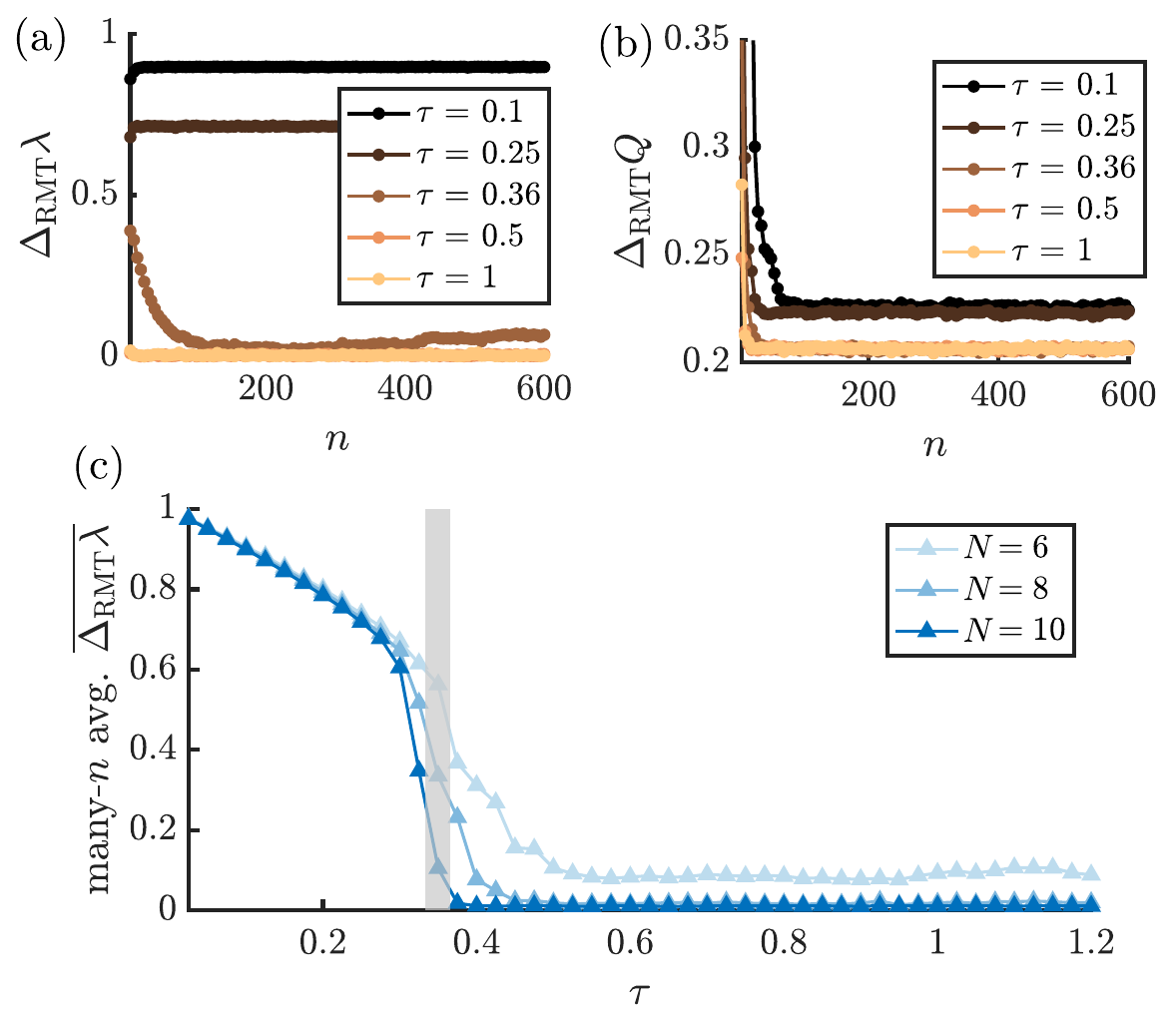}
	\caption{{\bf{Convergence of $\lambda_1$ and $Q$ to RMT prediction}}. Deviation of $\lambda_1$ (a) and $Q$ (b) from the respective RMT estimates, as a function of the number of cycles $n$, for different values of $\tau$ below (black and brown) and above (orange and yellow) the Trotterization threshold, for a system of $N=10$ spins. (c) Long-time average of $\Delta_{\mathrm{RMT}}\lambda$ as a function of $\tau$. The vertical shaded grey line marks the Trotterization threshold. Away from the high-frequency regime which decreases with system size. The results are obtained for the Trotterized XXZ/Heisenberg model with $J^{x}_j=0.9$, $J^{y}_j=1.5$, $J^{z}_j=0.5$, $B^{x}_j=0.65$, $B^{y}_j=0.3$, $B^{z}_j=0.45$. In these plots we set $n_{\mathrm{s}}\to\infty$, and use $1000$ initial states in random bases.}
	\label{fig:RMT_comparisons}
\end{figure}

In this appendix, we provide further analysis of the Trotterization threshold which can be detected via FHL. Additionally, we show how $\lambda_1$ [Eq.~\eqref{eq:lambda1}] can be used as an indicator of the onset of quantum chaos in Trotterized dynamics~\cite{Heyl2019,Sieberer2019,Kargi2021,Olsacher2022}. For large values of $\tau$ (in units of the Hamiltonian timescales) no approximately local effective Hamiltonian can be reconstructed, which is signaled by $\lambda_1$ saturating to an approximately constant value in $\tau$ after the scaling regime (see, e.g., Fig.~\ref{fig:Heisenberg}(a)). This corresponds to the transition to quantum chaos, where the properties of the operator $\hat{U}_{\tau}$ can be approximately described by random matrix theory (RMT)~\cite{Haake2018,DAlessio2014,Regnault2016,Kos2018}. This notion of chaos can be probed through $\lambda_1$, extracted from FHL, thanks to the following observation. When the dynamics are governed by unitaries from a given random matrix ensemble, and when the chosen initial states for the HL protocol consist of product states, we can analytically estimate the form of the squared constraint matrix $Q$, defined as
\begin{equation}
	Q=\frac{1}{N_{\mathrm{con}}}M^{\mathsf{T}}M \,\,,
\end{equation}
and thus the value of $\lambda_1$. Given these estimates, denoted with $Q_{\mathrm{RMT}}$ and $\lambda_{\mathrm{RMT}}$, respectively, we can compute the differences
\begin{align}
	&\Delta_{\mathrm{RMT}}\lambda = \frac{\big\vert\lambda_{\mathrm{RMT}}-\lambda_1/\sqrt{N_{\mathrm{con}}}\big\vert}{\lambda_{\mathrm{RMT}}} \,\,, \label{aeq:DeltaLambdaRMT} \\
	&\Delta_{\mathrm{RMT}}Q = \frac{\Vert Q_{\mathrm{RMT}}-Q\Vert}{\Vert Q_{\mathrm{RMT}}\Vert} \,\,, \label{aeq:DeltaQmatRMT}
\end{align} 
between them and the measured $Q$ and $\lambda_1$: a decrease of these differences after the threshold $\tau_{*}$ constitutes an \emph{indicator} of the onset of chaos. Here, $\Vert\cdot\Vert$ for a matrix denotes its Frobenius norm. 

For the details on the analytical derivations of $Q_{\mathrm{RMT}}$ and $\lambda_{\mathrm{RMT}}$ we refer the reader to our recent work~\cite{Olsacher2022} in the context of collective spin systems. The basic idea is to calculate the elements of $Q_{\mathrm{RMT}}$ as the expectation value of the elements of $Q$ over the relevant matrix ensemble, which is the CUE in the case of the XXZ/Heisenberg DQS in the main text, i.e.,
\begin{equation} \label{aeq:Q_expRME}
	(Q_{\mathrm{RMT}})_{j,k}=\frac{1}{N_{\mathrm{con}}}\sum_i\mathbbm{E}_{\hat{U}\in\mathrm{CUE}}\big[M_{i,j}M_{i,k}\big]\,\,,
\end{equation}
where $M_{i,j}=\mathrm{tr}(\hat{\rho}_i \hat{h}_j)-\mathrm{tr}(\hat{\rho}_i\hat{U}^{\dagger}\hat{h}_j\hat{U})\big)$ with $\hat{\rho}_i$ denoting the chosen initial states for the protocol. This can be done using the techniques developed in~\cite{Brower1996}. Then, $\lambda_{\mathrm{RMT}}$ can be estimated as the square root of the smallest eigenvalue of $Q_{\mathrm{RMT}}$.

Using these results, we compute the relative deviations \eqref{aeq:DeltaLambdaRMT} and \eqref{aeq:DeltaQmatRMT} for the XXZ/Heisenberg DQS described by the Trotter cycle in Eq.~\eqref{eq:HeisenbergTrotteriz}. The results are shown in Fig.~\ref{fig:RMT_comparisons}, for a zeroth order ansatz $\mathcal{A}_0$. Since in general one needs to perform a number $n$ of Trotter cycles which is large enough in order for $\hat{U}_{\tau}^n$ to resemble a random unitary~\cite{Sieberer2019}, we first study how $\Delta_{\mathrm{RMT}}\lambda$ and $\Delta_{\mathrm{RMT}}Q$ change as a function of the number of cycles $n$ ---or the total simulation time--- for fixed $\tau$, by repeating the learning procedure described in Sec.~\ref{sec:HLinDQS} with increasing value of final time. This is shown in Fig.~\ref{fig:RMT_comparisons}(a)-(b). For large values of $\tau$ (beyond the Trotterization threshold) these deviations decay to a small steady value, which is consistently smaller than the respective value for small $\tau$. Then, we compute the steady value of $\Delta_{\mathrm{RMT}}\lambda$ as a function of $\tau$, expecting it to be large for small $\tau$ and close to zero in the chaotic regime. This is confirmed by the results shown in Fig.~\ref{fig:RMT_comparisons}(c), where we see a sharp decrease of $\Delta_{\mathrm{RMT}}\lambda$ in correspondence of the Trotterization threshold. This analysis further corroborates the intuitive connection between the Trotterization threshold in DQS and the onset of quantum chaos in the properties of $\hat{U}_{\tau}$, identifying in $\lambda_1$ a measurable indicator of this transition.

\section{Estimate of the noise threshold in Hamiltonian and Liouvillian learning} \label{app:Elambda}

In this appendix, we derive the estimates for the cost functions $\lambda_1$ (for Hamiltonian learning) and $\Delta$ (for Liouvillian learning) in presence of measurement noise. We start with the derivation of the noise threshold in Eq~\eqref{eq:Elambda} for Hamiltonian learning. We denote the measured constraint matrix with $\widehat{M} = M + \epsilon E$, with $M$ being the exact constraint matrix (in absence of measurement noise), $E$ and additive error matrix and $\epsilon \sim n_{\rm s}^{-1/2}$. In the limit of large number of shots $n_{\rm s}$, we can approximate the error matrix elements $E_{i,j}$ as being independent and identically distributed normal random variables, i.e., $E_{i,j} \sim \mathcal{N}(0,1)$. In the following, we calculate the expectation value of the smallest singular value of $\widehat{M}$, that is $\lambda_1(\widehat{M})$, over the noise realizations of $E$. To this end, we consider the $N_{\mathcal{A}}\times N_{\mathcal{A}}$ squared constraint matrix 
\begin{equation}
	\widehat{Q} = \widehat{M}^{\dagger} \widehat{M} = Q + V(\epsilon) \,\,,
\end{equation}
where $Q=M^{\dagger}M$ and
\begin{equation}
    \label{aeq:Vpert}
	V(\epsilon) = \epsilon \, (M^{\dagger} E + E^{\dagger} M)+ \epsilon^2 E^{\dagger} E \,\,.
\end{equation}
From this definition, we have that
\begin{equation}
    \lambda_1(\widehat{M})=\sqrt{\varepsilon_1(\widehat{Q})} \,\,,
\end{equation}
where $\varepsilon_1(\widehat{Q})$ is the smallest eigenvalue of $\widehat{Q}$. We calculate $\varepsilon_1(\widehat{Q})$ using standard perturbation theory, taking $V(\epsilon)$ as a perturbation on the exact squared constraint matrix $Q$. We work under the assumption that $\lambda_1(M)\ll 1$, and hence $\varepsilon_1(Q)\ll 1$, which corresponds to the case of a complete ansatz, or an ansatz missing only small terms, e.g., the higher order terms in $\hat{H}^\mathrm{exp}_{\rm F}(\tau)$ for small $\tau$. Then, denoting with $\boldsymbol{v}_n$ the eigenvectors of $Q$ with eigenvalues $\varepsilon_n(Q)$, using standard perturbation theory up to second order in $\epsilon$ one has
\begin{equation}
    \varepsilon_1(\widehat{Q})=(\boldsymbol{v}_1|V(\epsilon)\boldsymbol{v}_1)-\sum_{n>1}\frac{|(\boldsymbol{v}_n|V(\epsilon)\boldsymbol{v}_1)|^2}{\varepsilon_n(Q)} + \mathcal{O}(\epsilon^3) \,\,,
\end{equation}
where $(\boldsymbol{v}|\boldsymbol{w})$ denotes the scalar product between two vectors. Using Eq.~\eqref{aeq:Vpert}, the identities
\begin{align}
	& \mathbb{E}_{E\sim\mathcal{N}(0,1)}\left[(\boldsymbol{v}|E^{\dagger}E\,\boldsymbol{w}) \right] = N_{\mathrm{con}}\,(\boldsymbol{v}|\boldsymbol{w}) \,\,, \\
	& \mathbb{E}_{E\sim\mathcal{N}(0,1)}\left[(\boldsymbol{v}|E^{\dagger}\boldsymbol{w})(\boldsymbol{w}|E\,\boldsymbol{v}) \right] = \Vert \boldsymbol{v} \Vert^2 \Vert \boldsymbol{w} \Vert^2 \,\,,
\end{align}
and keeping terms up to order $\epsilon^2$, we obtain
\begin{equation}
	\mathbb{E}_{E\sim\mathcal{N}(0,1)}[\varepsilon_1(\widehat{Q})] = \epsilon^2 (N_{\mathrm{con}} - N_{\mathcal{A}} + 1) + \mathcal{O}(\epsilon^3) \,\,.
\end{equation}
Using Jensen's inequality, we finally have
\begin{align}
	\mathbb{E}_{E\sim\mathcal{N}(0,1)}[\lambda_1(\widehat{M})] & \leq \sqrt{\mathbb{E}_{E\sim\mathcal{N}(0,1)}[\varepsilon_1(\widehat{Q})]} \\
	& = \epsilon \sqrt{ N_{\mathrm{con}} - N_{\mathcal{A}} + 1 } \,\,,
\end{align}
which is the equation in the main text.

We now show how this perturbative calculation can be extended to the case of Liouvillian learning explained in Sec.~\ref{sec:LiouvillianLearning}. As before, we denote with $\widehat{G} = G + \epsilon_G E$ the measured constraint matrix, with $G=(G^H\;G^{\mathcal{D}})$, and with $\widehat{\boldsymbol{b}}=\boldsymbol{b}+\epsilon\,\boldsymbol{e}$ the measured constraint vector, with $\boldsymbol{e}$ an error vector with elements being i.i.d.~normal random variables. Here $\epsilon \approx n_{\rm s}^{-1/2}$ while $\epsilon_{G}$ can be estimated by bounding the variance of the integrals \eqref{eq:GintH}-\eqref{eq:GintD} defining the matrix elements of $G$ as
\begin{equation}
    \epsilon_{G}^2\lesssim \frac{4n\tau^2}{n_{\rm s}}\approx 4n\tau^2\epsilon^2 \,\,,
\end{equation}
with $n$ being the number of Trotter steps used to evaluate the integral. The factor $4$ comes from the integrands in Eq.~\eqref{eq:GintD} being commutators of Pauli operators. Assuming $G^{\dagger}G$ being non-singular, we can write the solution of the noisy linear system $\widehat{G}\boldsymbol{c}=\widehat{\boldsymbol{b}}$ as
\begin{align}
    \boldsymbol{c}^{\rm rec}&=(G^{\dagger}G)^{-1}G^{\dagger}\big(\boldsymbol{b}+\epsilon(\boldsymbol{e}-2\tau\sqrt{n}E\boldsymbol{c}^{\rm rec})\big) \label{aeq:noisyLinSysSol} \\
    & = \boldsymbol{c}^{\rm ex}+\delta\boldsymbol{c} \,\,,
\end{align}
where $\boldsymbol{c}^{\rm ex}=(G^{\dagger}G)^{-1}G^{\dagger}\boldsymbol{b}$ is the exact coefficient vector. Assuming that a perturbative expansion in the small parameter $\epsilon$ is valid for $\delta\boldsymbol{c}$, i.e., $\delta\boldsymbol{c}=\sum_{m=1}^{\infty}\epsilon^m\boldsymbol{\delta}_m$, we can insert it in the r.h.s.~of Eq.~\eqref{aeq:noisyLinSysSol} obtaining
\begin{equation}
    \delta\boldsymbol{c} = \epsilon\,(G^{\dagger}G)^{-1}G^{\dagger}(\boldsymbol{e}-2\tau\sqrt{n}E\boldsymbol{c}^{\rm ex}) + \mathcal{O}(\epsilon^2) \,\,.
\end{equation}
The noisy residual vector $\boldsymbol{\Delta}=\widehat{G}\boldsymbol{c}^{\rm rec}-\widehat{\boldsymbol{b}}$ becomes thus 
\begin{equation}
    \boldsymbol{\Delta} = -\epsilon\,R(\boldsymbol{e}-2\tau\sqrt{n}E\boldsymbol{c}^{\rm ex}) + \mathcal{O}(\epsilon^2) \,\,,
\end{equation}
where $R=I-G\,(G^{\dagger}G)^{-1}G^{\dagger}=\sum_{\alpha|\lambda_{\alpha}(G)=0}|\boldsymbol{u}_{\alpha})(\boldsymbol{u}_{\alpha}|$, with $|\boldsymbol{u}_{\alpha})$ being the right singular vectors of $G$. Now we can calculate the expectation values over the random variables $\boldsymbol{e}$ and $E$ obtaining
\begin{equation}
    \mathbb{E}_{\boldsymbol{e},E\sim\mathcal{N}(0,1)}\Vert\boldsymbol{\Delta}\Vert\leq\epsilon\sqrt{(N_{\rm con}-N_{\mathcal{A}})(1+4n\tau^2\Vert\boldsymbol{c}^{\rm ex}\Vert^2)} \,\,.
\end{equation}
Note that this estimate depends on $\Vert\boldsymbol{c}^{\rm ex}\Vert$, which is in principle unknown. Denoting with $J$ the typical energy scales in the Liouvillian, $\Vert\boldsymbol{c}^{\rm ex}\Vert^2\sim J^2N$, hence the unknown term in the above equation is suppressed for $\tau\ll\tfrac{J^{-1}}{nN}$. When this is not the case, the above estimate is useful only if one has access to an estimate for $\Vert\boldsymbol{c}^{\rm ex}\Vert$.

\section{Details on Trotterization of 1D cluster Model}\label{app:1DCluster}

In this appendix we provide some additional details on the Trotterization of the 1D cluster model and on the application of FHL to this scenario. The Trotter sequence chosen in our simulations is $\hat{U}_{\tau}=\hat{U}^{XX}_{\tau}\,\hat{U}^{YX}_{\tau}\,\hat{U}^{XX}_{-\tau}\,\hat{U}^{YX}_{-\tau}$. Here, $\hat{U}^{XX}_{\tau}=\mathrm{e}^{-\mathrm{i}\tau J\,\hat{H}_{XX}}$ with $\hat{H}_{XX} = \sum_{j}\hat{\sigma}^x_{j}\hat{\sigma}^x_{j+1}$, while $\hat{U}^{YX}_{\tau}=\mathrm{e}^{-\mathrm{i}\tau\,\hat{\mathcal{H}}_{YX}(\tau)}$ with $\hat{\mathcal{H}}_{YX}(\tau)=J\hat{H}_{YX}+\mathcal{O}(\tau^2)=J\sum_{j}\hat{\sigma}^y_{j}\hat{\sigma}^x_{j+1}+\mathcal{O}(\tau^2)$. In order to derive the Floquet Hamiltonian of this Trotterization, we compute
\begin{align*}
	&\hat{U}^{XX}_{\pm\tau}\,\hat{U}^{YX}_{\pm\tau}=\mathrm{exp}\Big\{\mp\mathrm{i}\tau \Big(J\hat{H}_{XX}+J\hat{H}_{YX} \\
	&\quad\quad\quad\,\pm\tau J\sum_j\hat{\sigma}^x_{j}\hat{\sigma}^z_{j+1}\hat{\sigma}^x_{j+2}\pm\tau J\sum_{j<N}\hat{\sigma}^z_{j}+\mathcal{O}(\tau^2)\Big)\Big\} \,\,.
\end{align*}
Using this, we arrive at the Floquet Hamiltonian in Eq.~\eqref{eq:1DclusterFMHam}. 

In the simulations presented in Fig.~\ref{fig:ClusterModel}(a), we observe a $\lambda_1(\tau) \in\mathcal{O}(\tau)$ for an ansatz $\mathcal{A}_1$ (similarly for $\mathcal{A}_0\cup\mathcal{A}_1$). This seems to be in contrast with the expectation that, for an ansatz $\mathcal{A}_{0}\cup\mathcal{A}_{1}$, $\lambda_1(\tau)$ should scale as $\mathcal{O}(\tau^2)$. This unexpected behavior of $\lambda_1(\tau)$ is a consequence of the choice of the final times in the simulation of FHL, which we explain in the following. As specified in the caption of Fig.~\ref{fig:ClusterModel}, the final simulation times $T$ are chosen proportional to $\tau^{-1}$, since the couplings in the Floquet Hamiltonian of Eq.~\eqref{eq:1DclusterFMHam} are proportional to $\tau$, that is $J_{\mathrm{C}}=2J\tau$. This choice is made in order to keep the total `effective simulation time' $J_{\mathrm{C}}T$, and thus the (Frobenius) norm of the constraint matrix $M(\tau)$, approximately constant over the chosen range of $\tau$. To understand the reason why $\lambda_1(\tau)\in\mathcal{O}(\tau)$, we rewrite the elements of the constraint matrix as follows
\begin{align}
\begin{split}
	M_{i,j}(\tau)&=\langle \hat{h}_j \rangle_{i,0}-\langle \hat{h}_j \rangle_{i,T=n\tau} \\
	&=\langle \hat{h}_j \rangle_{i,0}-\langle \hat{h}_j \rangle^{\mathrm{target}}_{i,T}-\Delta_{i,j}(T,\tau) \\
	&\equiv M^{\mathrm{target}}_{i,j}-\Delta_{i,j}(T,\tau) \,\,.
\end{split}
\end{align}
In the second row above we denoted with $\langle \hat{h}_j \rangle^{\mathrm{target}}_{i,T}$ the expectation value of the ansatz operator $\hat{h}_j$ on the state $\ket{\psi_i}$ evolved with the target Hamiltonian $\hat{H}_{\mathrm{1DC}}$ of Eq.~\eqref{eq:1DclusterTarget} (with $J_{\mathrm{C}}=2J\tau$ and $B_{j<N}=2J\tau$), i.e., $\langle \hat{h}_j \rangle^{\mathrm{target}}_{i,T}=\langle\psi_i|\mathrm{e}^{\mathrm{i}\hat{H}_{\mathrm{1DC}}T}\hat{h}_j\,\mathrm{e}^{-\mathrm{i}\hat{H}_{\mathrm{1DC}}T}|\psi_i\rangle$, and $\Delta_{i,j}(T,\tau)$ denotes the deviation of the DQS expectation value from the corresponding target one due to Trotter errors. Such deviation scales as $\tau^2$, but also depends linearly on the total simulation time $T$: we can therefore write 
\begin{equation}
	\Delta_{i,j}(T,\tau)\in\mathcal{O}(T\tau^2)=\mathcal{O}(J\tau^{-1}\tau^2)=\mathcal{O}(\tau) \,\,.
\end{equation}
The DQS constraint matrix $M(\tau)$ can be written as the `target constraint matrix' $M^{\mathrm{target}}$, defined above, whose smallest singular value is exactly zero for an ansatz encompassing the target Hamiltonian as in Sec.~\ref{subsec:ClusterModel}, plus a perturbation whose norm scales linearly with $\tau$. As a consequence $\lambda_1(\tau)$ calculated from $M(\tau)$ also shows a linear behavior in $\tau$.

\section{Extension of the scaling relation to Floquet Liouvillian learning} \label{app:ScalingLiouvillian}
Here, we extend the scaling relation of Eq.~\eqref{eq:ScalingRelation} to the case of the Liouvillian learning discussed in the main text. Here, the role of $\lambda_1(\tau)$ as `learning error' is taken by $\Delta(\tau)=\Vert G(\tau)\,\boldsymbol{c}^{\mathrm{rec}}(\tau)-\boldsymbol{b}(\tau)\Vert$, with $G(\tau)=(G^H(\tau)\;G^{\mathcal{D}}(\tau))$ and $\boldsymbol{c}^{\mathrm{rec}}(\tau)=(\boldsymbol{c}^H_{\mathrm{rec}}(\tau)\;\boldsymbol{c}^{\mathcal{D}}_{\mathrm{rec}}(\tau))$ defined in Sec.~\ref{sec:LiouvillianLearning}. In order to show the relation
\begin{equation}
    \Delta(\tau)\in\mathcal{O}(\tau^K) \Leftrightarrow \Vert\mathcal{L}^\mathrm{exp}_{\mathrm{F}}(\tau)-\mathcal{L}_{\mathrm{rec}}(\tau)\Vert\in\mathcal{O}(\tau^K) \,\,,
    \label{aeq:scaling_liouv}
\end{equation}
we rewrite the components of the \emph{residuals vector} $\boldsymbol{\Delta}(\tau)=G(\tau)\,\boldsymbol{c}^{\mathrm{rec}}(\tau)-\boldsymbol{b}(\tau)$ using the superoperator formalism. Each constraint operator $\hat{A}_k$ or density matrix $\hat{\rho}(t)$ is represented by a vector $|A_k\rrangle$ or $|\rho(t)\rrangle$ in the Liouville space $\mathcal{H}\otimes\mathcal{H}$, and the Liouvillian $\mathcal{L}^\mathrm{exp}_{\mathrm{F}}(\tau)$ becomes a linear (super)operator acting on them. Denoting with $\llangle A_k|\rho(t)\rrangle=\mathrm{tr}\big(\hat{A}^{\dagger}_k\hat{\rho}(t)\big)$ the inner product in the Liouville space, we can write 
\begin{equation}
    \llangle A_k|\rho(n\tau)\rrangle-\llangle A_k|\rho(0)\rrangle=
    \int_0^{n\tau}\llangle A_k|\mathcal{L}^\mathrm{exp}_{\mathrm{F}}(\tau)|\rho(t)\rrangle\,\mathrm{d}t\,\,,
    \label{aeq:EherenfestLiouvilleSuperop}
\end{equation}
which holds for all constraints $\llangle A_k|$, initial states $|\rho(0)\rrangle$ and times $n\tau$. We can therefore write
\begin{align}
\begin{split}
    \Delta_k(\tau)&=\int_0^{n\tau}\llangle A_k|\mathcal{L}^\mathrm{exp}_{\mathrm{F}}(\tau)-\mathcal{L}_{\mathrm{rec}}(\tau)|\rho(t)\rrangle\,\mathrm{d}t \\
    &\equiv\int_0^{n\tau}\llangle A_k|\mathcal{V}(\tau)|\rho(t)\rrangle\,\mathrm{d}t \,\,,
\end{split}
\end{align}
which directly relates the components of the residuals vector with the (super)operator difference between the exact and the reconstructed Liouvillian. To show that \eqref{aeq:scaling_liouv} holds, it is sufficient to prove that for any $\hat{A}_k$ and $\hat{\rho}(t)$ being analytic functions of $\tau$ one has
\begin{equation}
    \mathcal{V}(\tau)\in\mathcal{O}(\tau^K) \Leftrightarrow \llangle A_k|\mathcal{V}(\tau)|\rho(t)\rrangle\in\mathcal{O}(\tau^K) \,\,,
\end{equation}
Direction ``$\Rightarrow$" is straightforward. Given that $\mathcal{V}(\tau)\in\mathcal{O}(\tau^K)$ element-wise, and for $|A_k\rrangle$ and $|\rho(t)\rrangle$ analytic functions of $\tau$, taking the expectation value $\llangle A_k|\mathcal{V}(\tau)|\rho(t)\rrangle$ is a linear operation that cannot lower the power $K$.\\
Let us now focus on direction ``$\Leftarrow$". We start by noticing that the constraint operators $|A_k\rrangle$ and the density operators $|\rho(t)\rrangle$ can be chosen as elements of an (hermitian) basis of the Liouville space. An example of such basis of density operators is given by the tensor product of bases of single-qubit density matrices, for instance as $\hat{\rho}_{l}\in\left\{\bigotimes_{j=1}^N\left\{\frac{\hat{\mathbbm{1}}\pm\hat{\sigma}^z_j}{2},\frac{\hat{\mathbbm{1}}-\hat{\sigma}^x_j}{2},\frac{\hat{\mathbbm{1}}-\hat{\sigma}^y_j}{2}\right\}\right\}$ with $l=1,\dots,\mathrm{dim}(\mathcal{H}\otimes\mathcal{H})$. Hence, since $\llangle A_k|\mathcal{V}(\tau)|\rho(t)\rrangle\in\mathcal{O}(\tau^K)$ for any $|A_k\rrangle$ and $|\rho(t)\rrangle$, we have that
\begin{equation*}
    \forall\,k,l\,,\; \big(\mathcal{V}(\tau)\big)_{k,l}=\llangle\rho_k|\mathcal{V}(\tau)|\rho_l\rrangle\in\mathcal{O}(\tau^K)\,\,,
\end{equation*}
meaning that $\mathcal{V}(\tau)\in\mathcal{O}(\tau^K)$ element-wise.

The observation that $\mathcal{V}(\tau)\in\mathcal{O}(\tau^K)\Leftrightarrow\Vert\mathcal{L}^\mathrm{exp}_{\mathrm{F}}(\tau)-\mathcal{L}_{\mathrm{rec}}(\tau)\Vert\in\mathcal{O}(\tau^K)$, with $\Vert\cdot\Vert$ denoting the Frobenius matrix norm, concludes the proof of \eqref{aeq:scaling_liouv}. As a final remark, we note that the integrand $\llangle A_k|\mathcal{L}^\mathrm{exp}_{\mathrm{F}}(\tau)|\rho(t)\rrangle$ in Eq.~\eqref{aeq:EherenfestLiouvilleSuperop} must be measured at integer multiples of $\tau$, which becomes the minimal time-step that can be used for evaluating the integral. Hence, the evaluation of the integral in Eq.~\eqref{aeq:EherenfestLiouvilleSuperop} comes with an inherent integration error due to the Trotterization, which scales as $\mathcal{O}(\tau^m)$ with an exponent $m$ controlled by the integration routine. Hence, in order to observe scaling $\Delta(\tau)\in\mathcal{O}(\tau^K)$ due to Trotter errors, one has to choose an integration routine with $m\geq K$.\\

\section{Numerical example of scalability} 
\label{app:scalability}

\begin{figure} [ht]
	\centering  
	\includegraphics[width=0.9\linewidth]{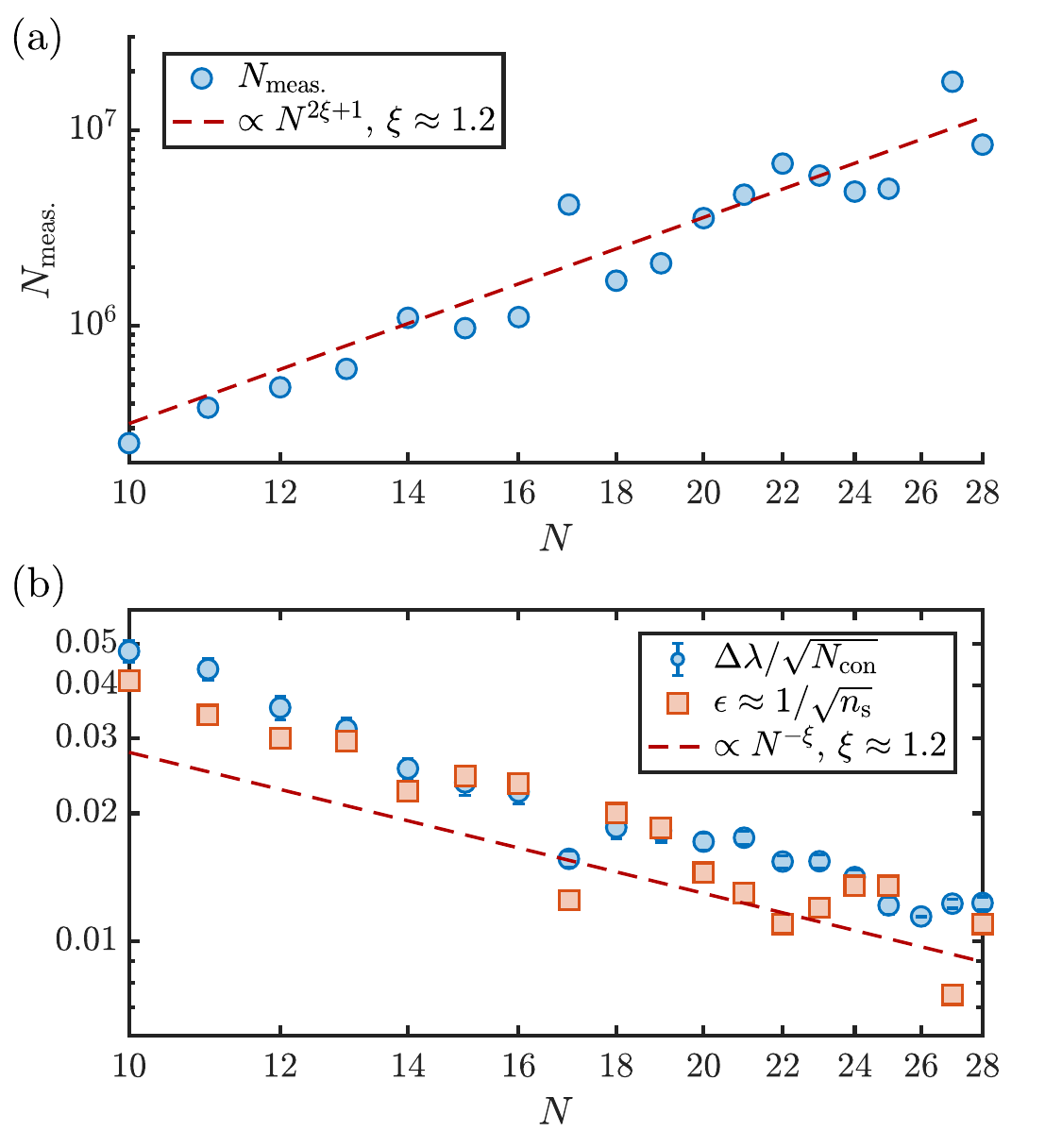}
	\caption{{\bf{Finite size scaling of measurement budget}}. (a) Total number $N_{\rm meas.}$ of measurement shots vs.~system size $N$, for learning the zeroth order terms in a DQS of a disordered Heisenberg/XXZ model, to an average accuracy of $\Vert\hat{\boldsymbol{c}}^{\rm rec}-\hat{\boldsymbol{c}}^{\rm ex}\Vert\approx 0.1$. The simulated data confirm the polynomial scaling of the measurement budget. (b) Spectral gap of the constraint matrix (blue data) and accuracy $\epsilon\approx 1/\sqrt{n_{\rm s}}$ per matrix element for achieving $\Vert\hat{\boldsymbol{c}}^{\rm rec}-\hat{\boldsymbol{c}}^{\rm ex}\Vert\approx 0.1$. The data are obtained using TEBD with $\tau=0.1$, evolving a set of $\approx 10 N$ deterministically chosen initial product states in the $z$ basis, up to a total time $T=15$. The maximal bond dimension was fixed to $3500$ to keep the truncation errors lower than $10^{-3}$ for the largest system sizes. The parameters used are $J^{x}_j=J^{y}_j=1+\delta_j^{x}$, $J^{z}_j=0.7+\delta_j^{z}$ with $\delta_j^{x}$ and $\delta_j^{z}$ randomly distributed in $[-0.1,0.1]$, and a disordered $z$ magnetic field $B^z_j\in[-3,3]$.}
	\label{fig:DMRGscaling}
\end{figure}

In this appendix, we provide a numerical indication of the scalability of our protocol to systems with large number of degrees of freedom, by showing the number of measurements required to achieve a fixed accuracy of the reconstructed Hamiltonian scales only polynomially in the system size $N$. Specifically, we simulated the reconstruction of the zeroth order terms in the Floquet Hamiltonian of the DQS of a disordered Heisenberg/XXZ model, for several values of $N$ and a fixed value of $\tau$ (below the Trotter threshold). To access larger system sizes, we used matrix product states (MPS) techniques implemented via the ITensor~\cite{Fishman2020} library. The Trotterized time-evolution is very naturally implemented on MPS using the time-evolving block decimation (TEBD) algorithm. In Fig.~\ref{fig:DMRGscaling} we show our results, where for each value of $N$ we added simulated measurements noise from a finite measurement budget $N_{\rm meas.}$ chosen to achieve a fixed parameter distance $\Vert\hat{\boldsymbol{c}}^{\rm rec}-\hat{\boldsymbol{c}}^{\rm ex}\Vert\approx 0.1$, with the hat denoting a normalized coefficient vector.

As it is visible from Fig.~\ref{fig:DMRGscaling}(a), the total number $N_{\rm meas.}$ of measurement shots required to reach a fixed accuracy in the reconstructed Hamiltonian scales polynomially with the system size $N$ (note the log-log scale). The exponent of this power-law scaling is governed by the locality of the underlying Hamiltonian ansatz, and by the scaling of the gap of the constraint matrix with $N$. More specifically, in the present case we observe a scaling $N_{\rm meas.}\sim N^{2\xi+1}$, where the `$+1$' contribution comes from the fact that the number of ansatz terms to be measured scales linearly with $N$, while the $2\xi$ contribution comes from the spectral gap of $M$, i.e., $\Delta\lambda=\lambda_2-\lambda_1$, as shown in the blue data of Fig.~\ref{fig:DMRGscaling}(b). There, we observe that the gap decreases as $\Delta\lambda\sim N^{-\xi}$ for the present example. To understand how the gap influences the number of measurements, we note that a perturbative result for the parameter distance $\Vert\hat{\boldsymbol{c}}^{\rm rec}-\hat{\boldsymbol{c}}^{\rm ex}\Vert$ in presence of measurement noise parametrized by $\epsilon\approx 1/\sqrt{n_{\rm s}}$ can be easily derived as $\Vert\hat{\boldsymbol{c}}^{\rm rec}-\hat{\boldsymbol{c}}^{\rm ex}\Vert\approx\frac{\epsilon}{\Delta\lambda}$~\cite{Bairey2019,Li2020}, which implies that the number of shots $n_{\rm s}$ to achieve a fixed parameter distance must scale as $\Delta\lambda^2$. Hence the observed scaling $N_{\rm meas.}\sim N^{2\xi+1}$. 

Finally, we note that the polynomial decay of $\Delta\lambda$ with the system size is a consequence of our choice of the model, which is here a disordered XXZ chain exhibiting MBL. For systems obeying the eigenstate thermalization hypothesis (ETH), the existence of a gap in the thermodynamic limit was proven in Ref.~\cite{Li2020}. However, as already mentioned, the polynomial decay of $\Delta\lambda$ only worsen the scaling of the measurement budget for HL, without invalidating its scalability.

\bibliography{refs.bib}

\end{document}